\documentclass[12pt]{article}

\usepackage{latexsym,epsfig,graphicx,amsmath,amssymb,amscd,undertilde}
\usepackage{multirow,psfrag,paralist,dsfont}

\usepackage[american]{babel}
\usepackage{epstopdf}
\usepackage{hyperref}

\usepackage{float}

\usepackage[sort, numbers]{natbib}

\newcommand{\epsilonhat}{\widehat{\epsilon}}

\newcommand{\alphahat}{\widehat{\alpha}}
\newcommand{\betahat}{\widehat{\beta}}
\newcommand{\tauhat}{\widehat{\tau}}
\newcommand{\gammahat}{\widehat{\gamma}}

\newcommand {\yhat}{\widehat{y}}

\newcommand{\N}{{\textrm{N}}}

\newcommand{\sigmahat}{\widehat{\sigma}}

\newcommand{\thetavec}{{\boldsymbol{\theta}}}

\newcommand{\thetavechat}{\widehat{\thetavec}}

\newcommand{\Cvec}{{\boldsymbol{C}}}

\newcommand{\Var}{{\rm Var}}

\newcommand{\var}{\textrm{Var}}
\newcommand{\Xvec}{\boldsymbol{X}}
\newcommand{\xvec}{\boldsymbol{x}}
\newcommand{\yvec}{{\boldsymbol{y}}}

\newcommand{\yeps}{y^{\epsilon}}
\newcommand{\yepshat}{\widehat{y}^{\epsilon}}

\newcommand{\yepsvec}{\boldsymbol{y}^{\epsilon}}

\begin{document}

\title{A Detailed Historical and Statistical Analysis of the Influence of Hardware Artifacts on SPEC Integer Benchmark Performance}

\author{
Yueyao~Wang$^1$, Samuel Furman$^2$, Nicolas Hardy$^2$, Margaret Ellis$^2$, \\ Godmar Back$^2$,  Yili Hong$^3$, and Kirk Cameron$^2$\\[1.5ex]
{\small $^1$School of Statistics and Mathematics, Zhejiang Gongshang University,}\\[-0.5ex]
{\small Hangzhou, Zhejiang, China 310018}\\
{\small $^2$Department of Computer Science, Virginia Tech, Blacksburg, VA 24061}\\
{\small $^3$Department of Statistics, Virginia Tech, Blacksburg, VA 24061}
}

\date{}

\maketitle

\begin{abstract}
The Standard Performance Evaluation Corporation (SPEC) CPU benchmark has been widely used as a measure of computing performance for decades. The SPEC is an industry-standardized, CPU-intensive benchmark suite and the collective data provide a proxy for the history of worldwide CPU and system performance. Past efforts have not provided or enabled answers to questions such as, how has the SPEC benchmark suite evolved empirically over time and what micro-architecture artifacts have had the most influence on performance?\textemdash have any micro-benchmarks within the suite had undue influence on the results and comparisons among the codes?\textemdash can the answers to these questions provide insights to the future of computer system performance? To answer these questions, we detail our historical and statistical analysis of specific hardware artifacts (clock frequencies, core counts, etc.) on the performance of the SPEC benchmarks since 1995. We discuss in detail several methods to normalize across benchmark evolutions. We perform both isolated and collective sensitivity analyses for various hardware artifacts and we identify one benchmark (libquantum) that had somewhat undue influence on performance outcomes. We also present the use of SPEC data to predict future performance.

\textbf{Keywords}: CPU Benchmark; Data Visualization; Libquantum; Regression Analysis, Sensitivity Analysis; SPEC Normalization.
\end{abstract}

\section{Introduction}\label{sec:introduction}

\subsection{Background and Motivation}

For more than 30 years, computing has pushed the boundaries of performance. From adherence to Moore's Law to the adoption and widespread use of high clock rates and parallelism in systems and micro-architecture, performance has continued to increase substantially. As performance has increased, so have the capabilities of our computing devices. From e-commerce to streaming services, we now watch videos on our pocket-sized devices in nearly any location while emergent immersible augmented reality technologies promise the demand for performance will not ebb for the foreseeable future.

The Standard Performance Evaluation Corporation (SPEC) CPU benchmark has been widely used as a measure of computing performance. The SPEC is an industry-standardized, CPU-intensive benchmark suite, stressing a system's processor, memory subsystem and compiler. The SPEC benchmark has grown and diversified since its inception in the late 1980's and the collective data is a proxy for the history of worldwide CPU and system performance.

While many have studied the intricacies and influence of the SPEC benchmarks, such studies tend to focus on historical perspective (\cite{specgrowth}), the likelihood of benchmark suites being representative of real workloads (\cite{spec2017broaden}), and similarity among benchmark suite codes (\cite{specsamebottlenecks}, \cite{specprogramsimilarity}, \cite{specredundancy}, and \cite{spec2006subsetting}).

To the best of our knowledge, none of these past efforts provide or enable answers to the following questions: a) How has the SPEC benchmark suite evolved empirically over time and what micro-architecture artifacts have had the most influence on performance? b) Have any micro-benchmarks within the suite had undue influence on the results and comparisons among the codes? and c) Can the answers to these questions provide insights to the future of computer system performance?

In our survey of the prevailing literature, the closest work to ours is the Stanford CPU database (CPUdb) \cite{stanford2012}. In their work, the authors assembled CPU information from a variety of sources and created an online repository for researchers to explore. A number of analysis and visualization tools resulted from this effort. However, while the database captures CPU details going back to the 1970's, the performance studies only go back to the 2006 data. In our work, we not only extend our analyses to pre-2006 data, we also leverage a lineage database \cite{sam2021thesis} that builds off the CPUdb and other sources and adds entity relationships that capture the evolution of microprocessors over time. With this additional information, we can conduct more detailed statistical studies of the implications of hardware artifacts on performance outcomes dating back to the late 1990s.

In this paper, we detail our historical and statistical analysis of specific hardware artifacts (clock frequencies, core counts, etc.) on the performance of the SPEC benchmarks since 1995. We discuss several methods in detail to normalize across benchmark iterations. We perform both isolated and collective sensitivity analyses for various hardware artifacts and we identify one benchmark (libquantum) that had somewhat undue influence on performance outcomes. We also discuss the use of SPEC data to predict future performance. Due to page limit, we chose one of the SPEC benchmarks, the SPEC base integer speed, to present the analysis results because it is widely used. However, our scope is not limited to the SPEC integer benchmark. The analysis framework can be inherited in other benchmarks on the CPU and system level to understand the evolution trend.

\subsection{Literature Review and Related Work}

As an important benchmark in the microprocessor industry, many efforts have been made to better understand the SPEC CPU benchmarks \cite{speccpu}. Of particular interest is the influence of various hardware components on performance outcomes. As the main metric for performance in this study, it is important that we understand the history of the SPEC CPU benchmarks in more detail. The information on SPEC design and how the SPEC benchmark changed over time can be found in \cite{specgrowth}, \cite{spec2017broaden}, and \cite{evolandevalspec}. Each SPEC benchmark release (SPEC95, SPEC2000, SPEC2006, SPEC2017) contains
multiple suites of microbenchmarks that provide overall benchmark scores~
\cite{speccpu}. Some details on the SPEC workload characteristics subsetting (e.g., instruction counts, memory footprints) are available in \cite{specsamebottlenecks}, \cite{specprogramsimilarity}, \cite{specredundancy}, and \cite{spec2006subsetting}. Regarding the previous data collection efforts, Danowitz et al.~\cite{stanford2012} used SPEC to discuss a wide variety of CPU performance metrics over time. Furman~\cite{sam2021thesis} used SPEC to track the performance of processors that follow particular lineages. Hardy~\cite{nicolas2021thesis} provided further details on gathering SPEC CPU information and discussed some differences between SPEC and other benchmarks.

Compared to \cite{stanford2012}, this work differentiates and focuses more on in-depth analyses (e.g., expanding on libquantum and on normalization methodology).
The major contributions of this paper are as follows.

\begin{inparaitem}
\item We first compare two approaches, the regression approach and the constant approach, to normalize the benchmark score across different SPECs. Although the $R^2$ of the regression method to normalize the score is slightly higher for some situations, we found that the constant approach is much simpler.

\item Then a sensitivity analysis of system factors and the overall benchmark score is conducted to understand how the systems' configuration and hardware factors impact performance qualitatively and quantitatively. These allow us to understand the impact of system factors such as core counts, frequency, and auto-parallel.

\item We also conduct a deeper analysis of the libquantum microbenchmark and provide insights into why it was discontinued after 2006. The analysis reveals that libquantum has a tremendous influence on the benchmark; no other microbenchmarks are comparable.

\item We build statistical models to predict the performance score in the future and quantify the associated uncertainty. We provide statistical inference on both the mean trend as well as the individual computer's performance in the future.

\end{inparaitem}

\subsection{Overview}
The rest of this paper is organized as follows. Section~\ref{sec:method} introduces the methodology we use to construct the datasets and analysis. Section~\ref{sec:results} presents the results. Section~\ref{sec:prediction} contains the prediction of future performance, and Section~\ref{conclusion} provides concluding remarks.

\section{Methodology}\label{sec:method}
\subsection{SPEC Data Collection}

To conduct our analysis, we pull from the CSGenome project's (\texttt{csgenome.org/about}) key artifact, the iLORE database~\cite{sam2021thesis}. iLORE is the first known attempt to use publicly available data to capture a lineage of computer system performance over time. iLORE is constructed using computer system, benchmark, and computer component attribute data collected using Python scripts that leverage the popular Beautiful Soup~\cite{bs4} and Pandas libraries~\cite{Pandas} in concert with manual efforts.

We refer to records that integrate measures of computer performance with system specifications as computer system and benchmark data.
A handful of organizations maintain repositories of computer system and benchmark data (\cite{speccpu} and \cite{top500lists}). These repositories target different market segments of computing and contain varied attributes depending on the goals and focus of the maintaining organization. To illustrate, TOP500~\cite{top500lists} and Green500~\cite{green500lists} lists target the world's fastest super computers while the SPEC CPU suites focus on individual enterprise machines.

Hardware manufacturers and independent curators maintain publicly available repositories of computer component attribute data. We gather processor attribute data from three key sources. The first two are online repositories of product specifications maintained by the microprocessor manufacturers Intel and AMD (\cite{amdprocessorspecs}, and \cite{arkintel}). The final source is a previous effort made by \cite{stanford2012} to construct a database of microprocessors that is useful for investigating trends in the history of microprocessors \cite{stanfordcpudbwebsite}.

TOP500 and Green500 publish their biannual lists on their websites in several downloadable formats including XML and EXCEL files.
SPEC provides full access to all results for suites dating back to SPEC CPU95. SPECCPU data is available as HTML, CSV, and text files.
Intel maintains a web page for each modern processor that contains commonly-tracked attributes such as clock frequencies and core counts and supplemental information including integrated graphics details.
AMD provides data on their processor offerings from the past five years. The attributes maintained here are comparable to Intel's repository. AMD offers its data in bulk in the form of a CSV file.
Danowitz et al.~\cite{stanford2012} expose their CPU DB as a collection of CSV files.

Data collection across all sources occurs in two steps. First, we download the relevant files from each source to a local storage device. Top500, Green500, AMD, and the CPU DB natively support downloading and creating a copy. Data from Intel and SPEC is stored in a hierarchical structure across many pages. Starting from SPEC and Intel's root pages, we crawl benchmark and product specifications and download the HTML source file of each processor page to disk.
Second, we extract relevant information from local copies using the xlrd, BeautifulSoup, and Pandas Python libraries.

Newly extracted data is not yet amenable to insertion into the iLORE data model. We use a pipe-and-filter architecture to resolve data issues within and across sources. Embedded attributes represent a class of intra-source data issues. Before populating the iLORE data model, we cleaned the systems and benchmarking data with the computer component data. The goal is to accurately identify a full processor record among several thousand that matches with the incomplete set of processor information included in each benchmark record. After preparing the component and benchmark data and establishing our data model, we populate the database. To ensure the presence of the processor records to which benchmark data were manually linked, we build the processor tables before building the systems and benchmark tables.

From the systems dump table, we use SQLAlchemy to normalize categorical variables like country, application, and interconnect, to their look-up tables. Next, we extract information for individual benchmarks. Common data like information sources are aggregated in a joint benchmark table that links to individual benchmarks. Third, we construct the system table using both the system dump and look-up tables. Finally, we link together each benchmark with a system and system hardware configuration.

\subsection{Data Summary}
We focus on integer benchmarks in this paper. Integer performance has been of interest to the research community and industry. Table~\ref{tab:microbenchmarks.evolution} visualizes how the microbenchmarks change over the years for the integer benchmark. Table~\ref{tab:sys.factor.summary} provides the mean values of system factors of records in SPEC suites. We can see that the average number of cores increases along the SPEC suites. The increase in frequency is obvious between SPEC 1995 and 2000. However, after SPEC 2006, the averaged frequency is almost unchanged. L3 cache size increases 2.4 times between SPEC 1995 and 2000 and around 1.5 times between later SPEC suites. The average threads per core are around 1 at SPEC 1995 and gradually increase to around 2 at SPEC 2017. Table~\ref{tab:score.summary} summarizes the values of base integer speed and rate across SPEC suites. Table~\ref{tab:score.summary} shows that the ranges of scores vary widely for each SPEC suite. For example, the base integer speed score range is $[1.06, 46.6]$ at SPEC 1995 but changes to $[93.7, 3108]$ at SPEC 2000, indicating that scores across SPEC suites can not be compared directly. Therefore, it is necessary to normalize scores across SPEC suites to make the scores comparable.

\begin{table}
\caption{SPEC2017 integer programs and the evolution of the SPEC benchmarks over time. The benchmark descriptions on the left are for SPEC2017 only and do not apply to earlier versions. Programs in the same row from different generations of SPEC are generally not related. Here, ``XML to HTML'' means XML to HTML conversion via XSLT, ``AI1'' means alpha-beta tree search (chess), ``AI2'' means Monte Carlo tree search (Go), and ``AI3'' means recursive solution generator (Sudoku).}
\label{tab:microbenchmarks.evolution}

\begin{center}
{\scriptsize
\begin{tabular}{c|c|c|c|c}
\hline
\hline
     &\multicolumn{4}{c}{SPEC} \\\cline{2-5}
     & 2017 & 2006 & 2000 & 1995 \\
\hline
GNU C Compiler &\multicolumn{3}{c}{$\longleftarrow$}&gcc\\
\hline
Perl Interpreter &\multicolumn{3}{c}{$\longleftarrow$}&perl\\
\hline
Route Planning &\multicolumn{2}{c}{$\longleftarrow$}&mfc&\\
\hline
General Data Compression        &XZ&&bzip2&\\
Discrete Event Simulation       &$\longleftarrow$&omnetpp&vortex&\\
XML to HTML &$\longleftarrow$&xalancbmk&gzip&\\
\hline
Video Compression &X264&h264ref&eon&\\
AI1      & deepsjeng &xjeng&twolf&\\
AI2      & leela &gobmk&vortex&\\
AI3      & exchange2 &astar&vtr&\\
 &&hmmer& carfty&\\
 &&libquantum&parser&\\
\hline
\hline
\end{tabular}
}
\end{center}
\end{table}

\begin{table}
\caption{The average values of system factors across SPEC suites. ``ATPC'' means average threads per core.}\label{tab:sys.factor.summary}

\begin{center}
\begin{tabular}{rrrrr}
  \hline
  \hline
  \multirow{2}{*}{SPEC} & average  & average  & average  & \multirow{2}{*}{ATPC} \\
  & cores & freq (MHz) & cache (kB) & \\ \hline
 1995 & 1.00 & 319.25 & 4840.73 & 1.02   \\
 2000 & 2.44 & 2219.51 & 11777.41 & 1.19  \\
 2006 & 8.30 & 2519.37 & 19422.78 & 1.72 \\
 2017 & 16.49 & 2571.18 & 30763.64 & 1.94  \\
   \hline
   \hline
\end{tabular}
\end{center}
\end{table}

\begin{table}
\caption{Summary of base integer speed and rate across SPEC suites.}\label{tab:score.summary}

\begin{center}
\begin{tabular}{r|rrrr}
  \hline
  \hline
\multirow{2}{*}{SPEC}    & \multicolumn{4}{c}{Base Integer Speed} \\
\cline{2-5}
   &  max  & average  & min  & \# records  \\
  \hline
 1995  & 46.60   &   12.16 &  1.06 & 548   \\
 2000  & 3108.00 & 1320.62 & 93.70 & 1374  \\
 2006  & 84.10   &   44.67 &  1.00 & 9285  \\
 2017  & 12.40   &    8.93 &  1.00 & 4144  \\\hline & \multicolumn{4}{c}{Base Integer Rate}\\\hline
 1995& 17239.00 & 750.25 & 13.30 & 773  \\
 2000& 4236.00 & 84.38 & 2.30 & 2481    \\
 2006& 50200.00 & 693.35 & 1.00 & 15039 \\
 2017& 7100.00 & 222.08 & 1.00 & 5376   \\
   \hline
   \hline
\end{tabular}
\end{center}
\end{table}

\subsection{Normalization of SPEC Score Across Years}

Understanding the evolution of processor performance requires a mechanism to compare across the SPEC benchmark releases: SPEC95, SPEC2000, SPEC2006, and SPEC2017. The adoption of a new release  of SPEC benchmark suites can take some time, thus there is a transition period when, for example, both SPEC2000 and SPEC2006 are in use. Danowitz et al.~\cite{stanford2012} used the geometric mean of score ratios between the overlap sets of two SPEC suites as the conversion factor. The early SPEC suite score is converted to the new suite by multiplying the conversion factor. In this study, we also tried to use linear regression models to compute the score ratio. After obtaining the overlap set between two SPEC suites, we predict the log of the score ratio as a linear combination of the log old scores as well as a set of system factors, such as frequency, the number of cores, and threads per core, which is represented as,
$$\log\left(\frac{\text{score}_{new}}{\text{score}_{old}}\right) = \beta_0 + \beta_1 \log(\text{score}_{old}) + \beta_2\,\text{system factor}.$$
For different SPEC suites, we may include different sets of system factors in the linear model.

\begin{figure}
\begin{center}
\begin{tabular}{c}
\hspace{2ex}
\includegraphics[width=0.9\linewidth]{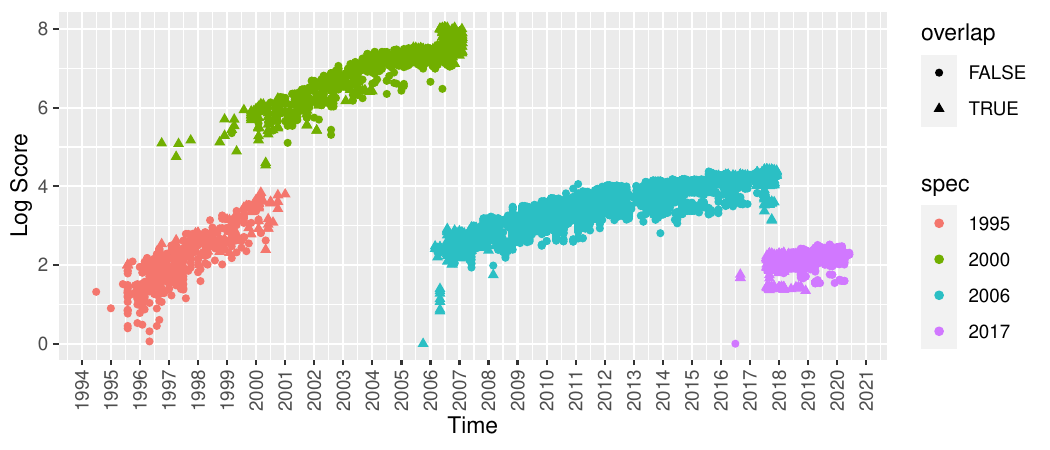}\\[-1ex]
{(a) Original}\\[2ex]
\includegraphics[width=0.9\linewidth]{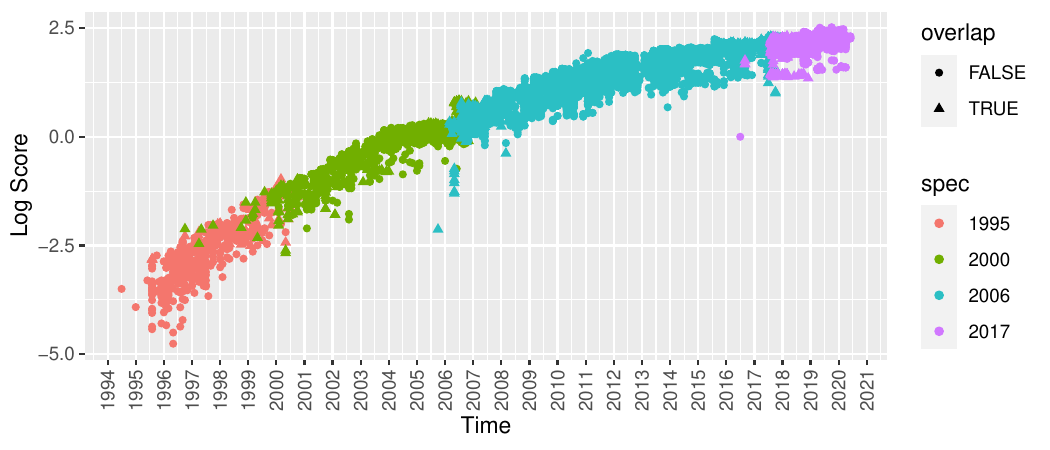}\\[-1ex]
 {(b) Constant}\\[2ex]
\includegraphics[width=0.9\linewidth]{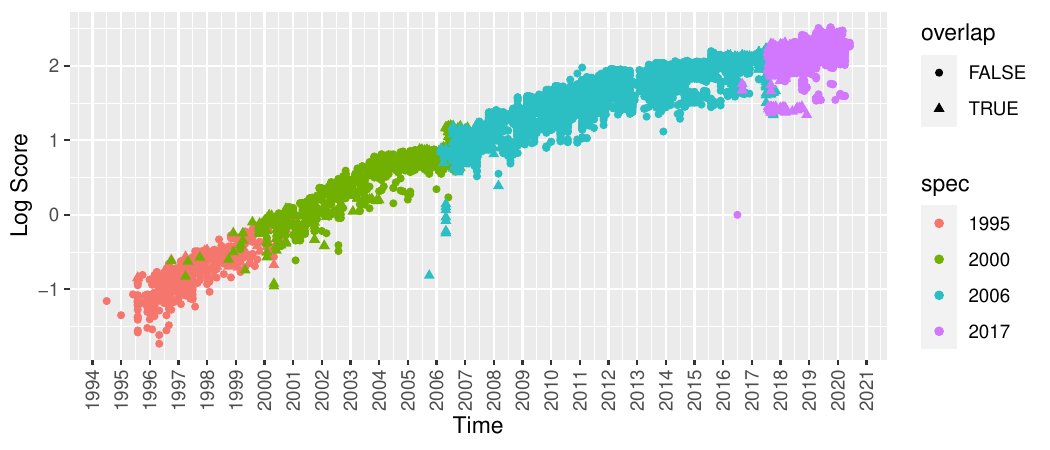}\\[-1ex]
{(c) Regression}
\end{tabular}
\caption{Two approaches to normalize base integer speed across years. The resolution of the time scale is one month and the time origin is set to 1995-08-01.}\label{fig:integer.speed.norm2}
\end{center}
\end{figure}

For these two approaches, we use the converted new score to predict the true new score. For example, the $R^2$ computed from cross-validation is summarized in Table~\ref{tbl:norm.r2.speed} for base integer speed, which we see that the linear regression approach has higher $R^2$. This finding indicates that when updating a benchmark to the next generation, making use of the previous suite score can be informative for normalization. Figure~\ref{fig:integer.speed.norm2} also shows the two approaches to normalize base integer speed across years. However, such improvement is relatively small compared to the effort we need to build the linear regression model. Therefore, for the normalization, we choose to use the approach proposed in \cite{stanford2012} for the following analysis in this paper.

\begin{table}
\caption{The $R^2$ of two types of conversion for base integer speed.}
\label{tbl:norm.r2.speed}

\begin{center}
\begin{tabular}{lrr}
\hline
\hline
		SPEC Suites		&  Regression &  Constant  \\
\hline
  1995-2000 & 0.889 & 0.782 \\
  2000-2006 & 0.738 & 0.697 \\
  2006-2017 & 0.802 & 0.825 \\
				\hline
				\hline
\end{tabular}
\end{center}
\end{table}

\subsection{Methods for Statistical Analysis}

The approaches that we used for sensitivity analysis fall within two categories, graphical visualization and quantitative modeling. For the visualization approach, we design a graph that can simultaneously visualize the effect of four variables on the benchmark scores. In particular, we did sensitivity analysis of time, core count, L3 cache size, frequency, transistor count, and auto-parallel (i.e., compiler flags for either automatic or explicit parallelism \cite{nicolas2021thesis}). Using graphs, we can see how the time trends of benchmark scores differ from different variables.

The second approach is based on statistical regression models. The regression approach provides a quantitative description of the relationship between benchmark scores and system variables, beyond the visualization approach. We consider both linear regression and nonlinear regression, depending on the specific functional relationship. Using the regression, we can see the quantitative effects of different variables on the benchmark scores, through the estimated regression model parameters.

With the comprehensive SPEC data in the CSGenome database and the use of statistical tools, we are able to do the following analysis for the integer benchmark performance.

\begin{inparaitem}
\item We study the relationship between the overall scores and microbenchmark scores.

\item We study the effect of system factors on benchmark scores.

\item We perform sensitivity analyses of microbenchmark scores and identify reasons why the libquantum microbenchmark is particularly influential.

\item We perform analyses to separate the effect of system factors.

\end{inparaitem}
\noindent
The detailed results and findings are presented in Section~\ref{sec:results}.

\section{Data Analysis Results}\label{sec:results}

\subsection{Overall Score and Microbenchmarks}

To understand the relationship between microbenchmarks and the overall score (i.e., base integer speed), first, we want to understand how the microbenchmarks contribute to the overall score computing for each SPEC suite.

According to the SPEC documentation, after the benchmarks are run on the system under test (SUT), a ratio for each of them is calculated using the run time on the SUT and a SPEC-determined reference time. The base integer speed (rate) is computed by the geometric mean of normalized (throughput) ratios when the benchmarks are compiled with base tuning. Suppose we have $p$ microbenchmarks in a SPEC suite, then the calculation can be expressed as,
\begin{align}\label{eqn:base.int.score}
\text{Base integer score} = &\left(\frac{\text{microbenchmark}_1}{\text{reference time}_1}\right)^{\frac{1}{p}} \times \dots \times  \left(\frac{\text{microbenchmark}_{p}}{\text{reference time}_{p}}\right)^{\frac{1}{p}}.
\end{align}
If we take log on both sides of \eqref{eqn:base.int.score}, we can obtain,
\begin{align}\label{eqn:log.base.int.score}
\log\left({\text{Base integer score}}\right) = \frac{1}{p}\sum_{i=1}^{p}\log(\text{microbenchmark}_i)+ c,
\end{align}
where $c$ is constant. Note that equation \eqref{eqn:log.base.int.score} indicates that the $\log\left({\text{Base integer score}}\right)$ and the $\log\left(\text{microbenchmarks}\right)$ should be linearly correlated for each SPEC suite. We include the visualization of base integer speed versus gcc and perl for each SPEC suite in Figure~\ref{fig:overall_micro_speed}.

\begin{figure}
    \centering
\begin{tabular}{c}
\includegraphics[width = \linewidth]{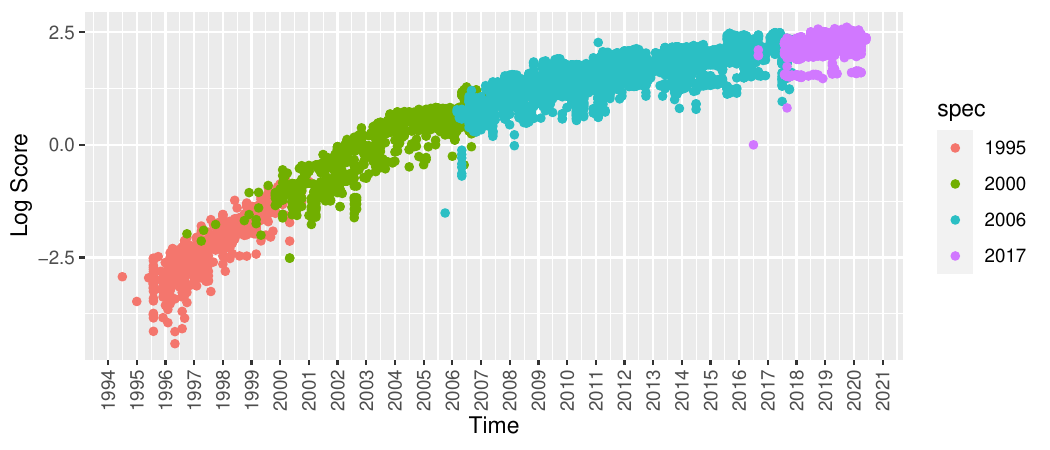}\\[-1ex]
  (a) GCC \\[-0.5ex]
\includegraphics[width = \linewidth]{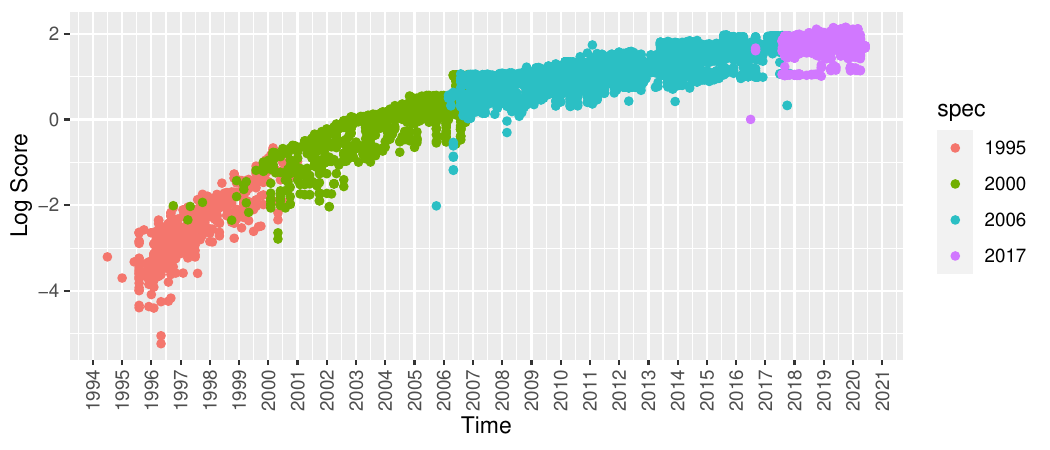}\\[-1ex]
(b) Perl
\end{tabular}
\vspace{-3ex}
    \caption{The normalized gcc and perl benchmarks for speed.}
    \label{fig:gcc.perl.speed.normalize}
\end{figure}

\begin{figure}
    \centering
\begin{tabular}{cc}
\includegraphics[width = 0.33\linewidth]{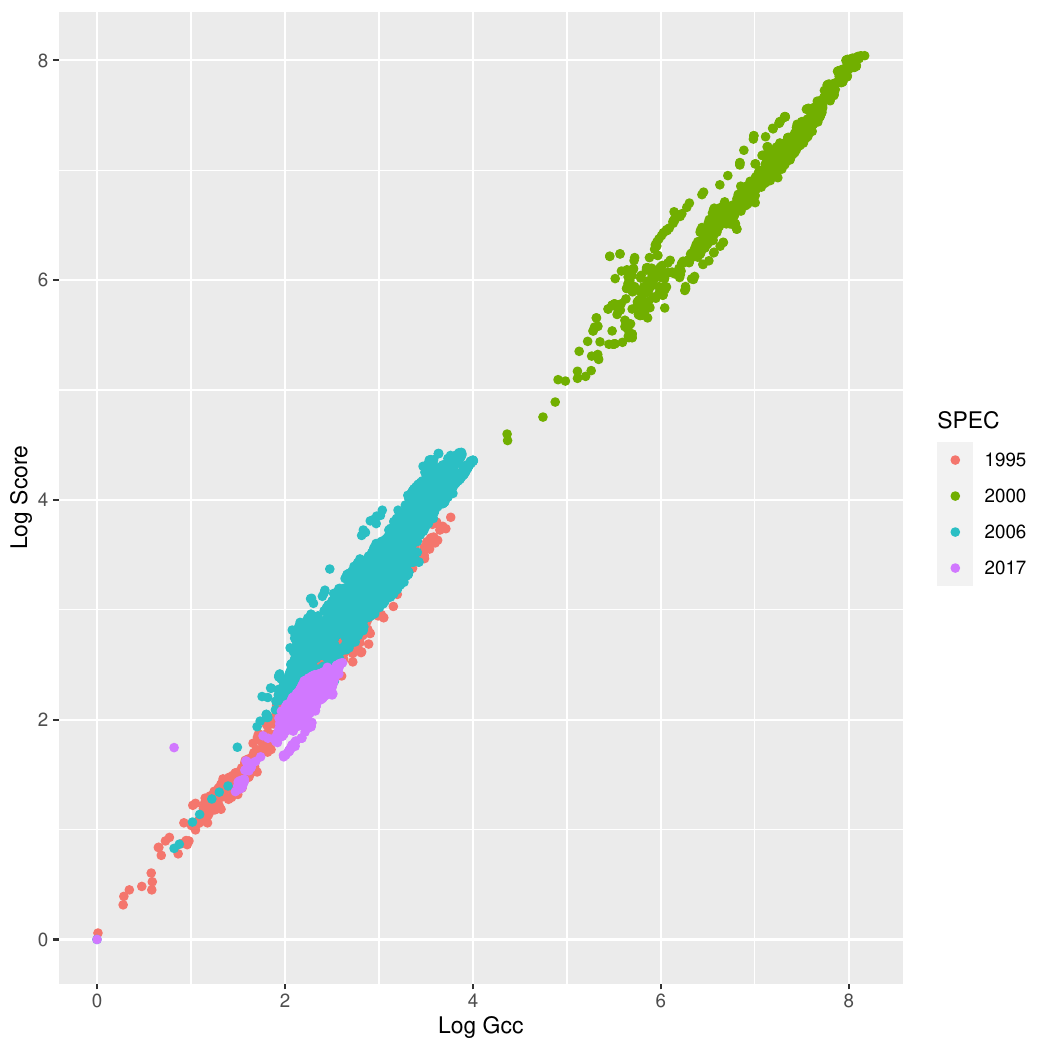} &
   \includegraphics[width = 0.33\linewidth]{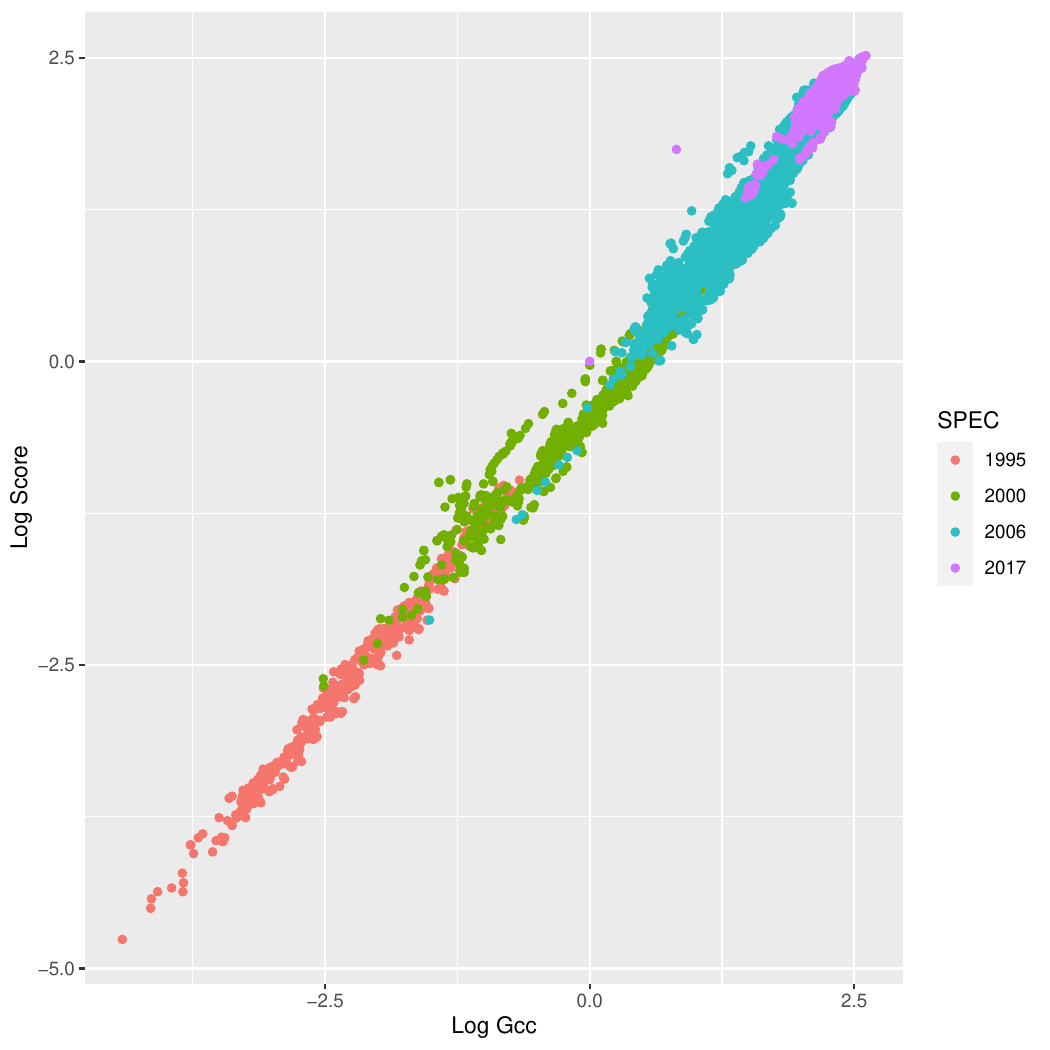} \\[-1ex]
   (a) GCC w/o normalization & (b) GCC with normalization\\[-0.5ex]
   \includegraphics[width = 0.33\linewidth]{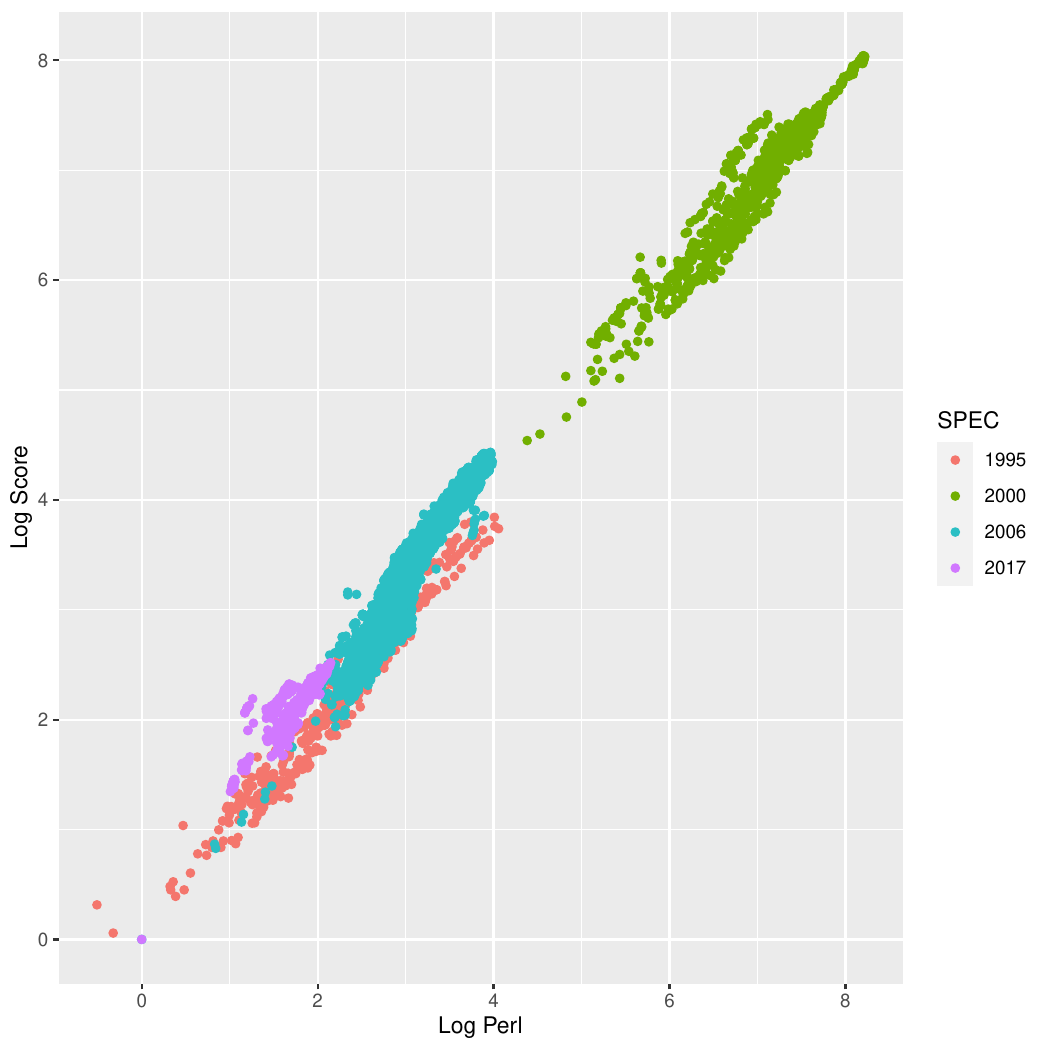} &
   \includegraphics[width = 0.33\linewidth]{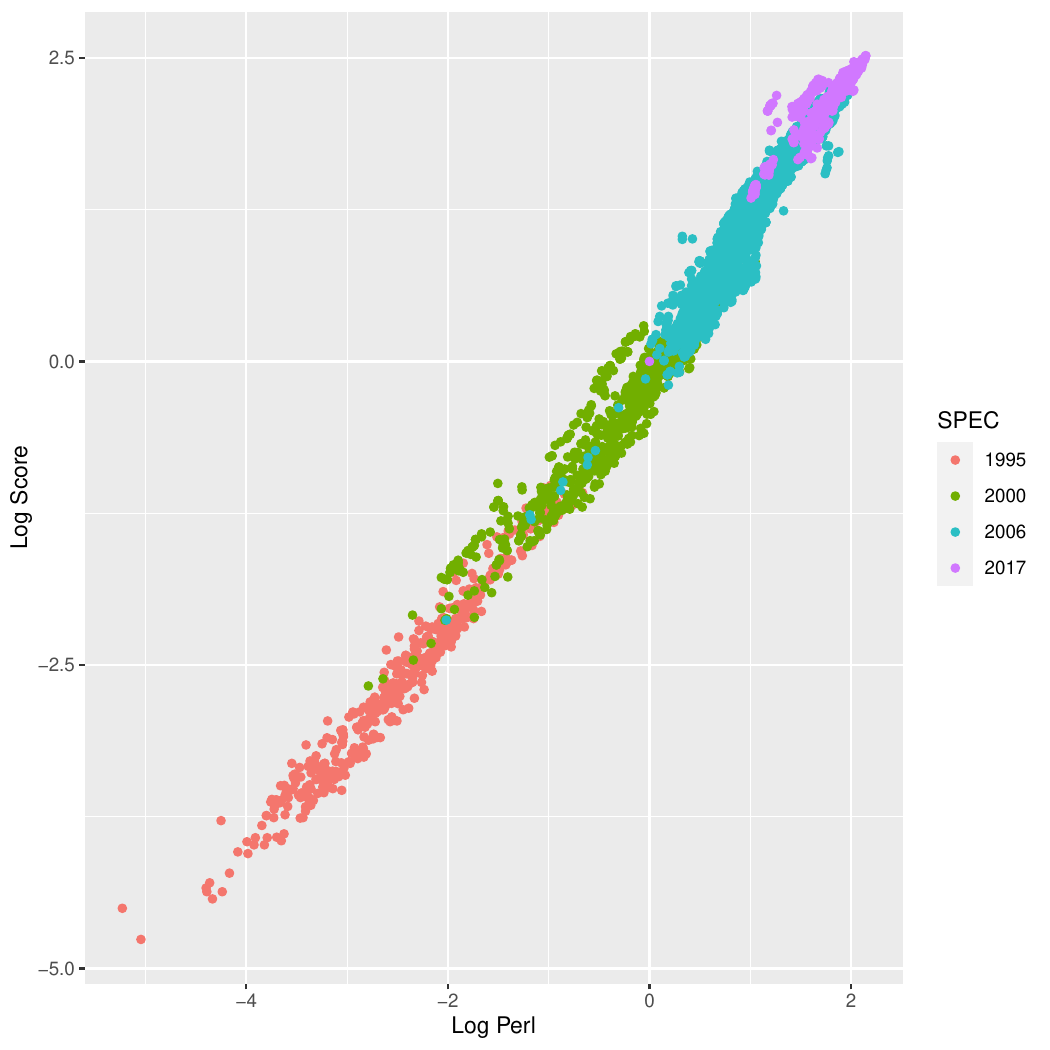}\\[-1ex]
(c) Perl w/o normalization & (d) Perl with normalization
\end{tabular}
    \caption{The normalized gcc and perl benchmarks versus normalized base integer speed.}
    \label{fig:gcc.perl.vs.speed}
\end{figure}

With the normalization we discussed above, we are able to analyze the microbenchmark effects on the overall score across the SPEC suites, instead of focusing on a single SPEC suite. However, as presented in Table~\ref{tab:microbenchmarks.evolution}, not all microbenchmarks survive during the evolution. Only gcc and perl exist for all suites. Therefore, we normalized these two microbenchmarks using the same methodology for the overall score and visualized their relationship with base integer speed.
Figure~\ref{fig:gcc.perl.speed.normalize} shows the normalization of gcc and perl benchmarks for  base integer speed calculation across years. Figure~\ref{fig:gcc.perl.vs.speed} (on page 6 for double-column figure) presents the gcc and perl versus base integer speed before and after normalization. We can see that normalization makes the micro and overall benchmarks follow a straight line.

\subsection{System Factors}\label{sec:sensitivity.sys}

\begin{figure}
    \begin{center}
    \includegraphics[width = \linewidth]{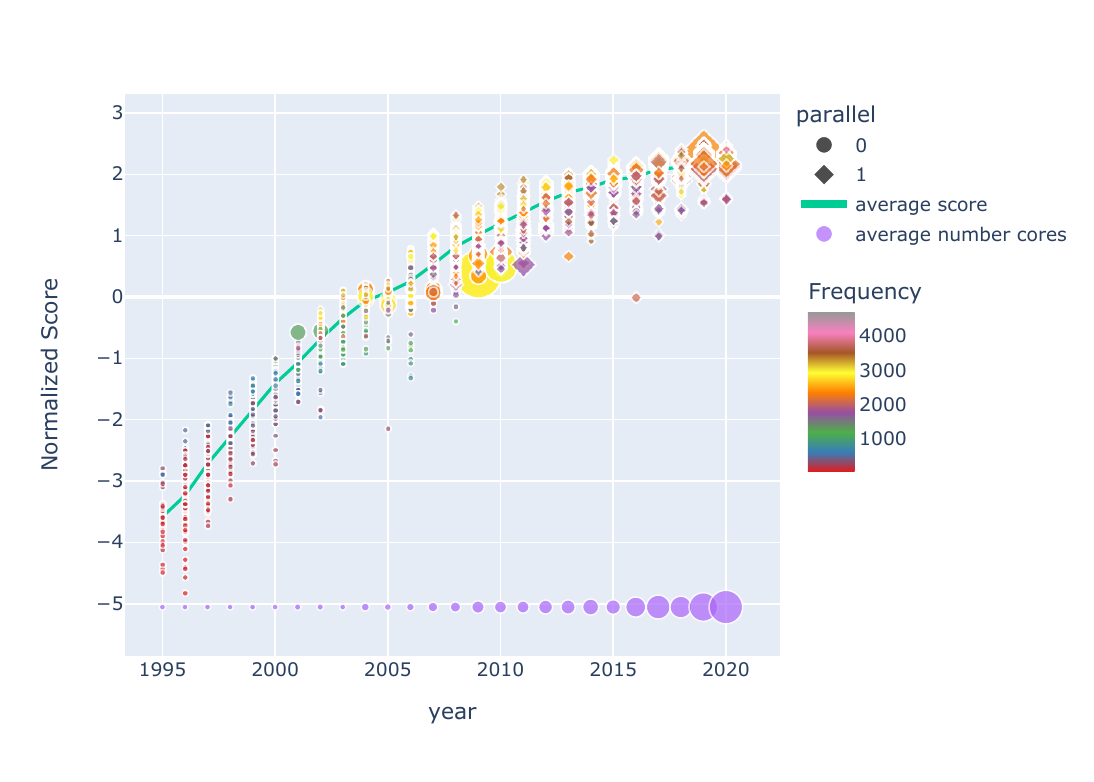}
    \vspace{-5ex}
    \caption{Normalized base integer speed score across years.}
    \label{fig:sen_speed}
    \end{center}
\end{figure}

System factors such as core counts, clock speed, and transistor counts often play important roles in computer performance. With the normalization method, we are able to  analyze the impact of system factors on the normalized score across years. Figure~\ref{fig:sen_speed} visualizes the impacts of frequency, core counts, and auto parallel to the normalized base scores across years. For base integer scores, we select all machines with base integer speed scores and normalize their scores based on the constant conversion mentioned in Section~\ref{sec:method}. For each machine record, we visualize their scores versus year in Figure~\ref{fig:sen_speed}, in which we use dot size to show the core counts of the machine, dot color to present the machine's frequency, and dot shape to visualize whether the machine has auto parallel. For base integer speed in Figure~\ref{fig:sen_speed}, we can see that as time passes, machines with auto parallel tend to gain higher scores. Both frequency and core counts increase as time passes. For more recent years, machines with higher core counts tend to have higher scores while the contribution of frequency is less influential. Both high and low score machines can have a high frequency.

\begin{figure}
    \centering
\begin{tabular}{cc}
 \includegraphics[width = 0.3\linewidth]{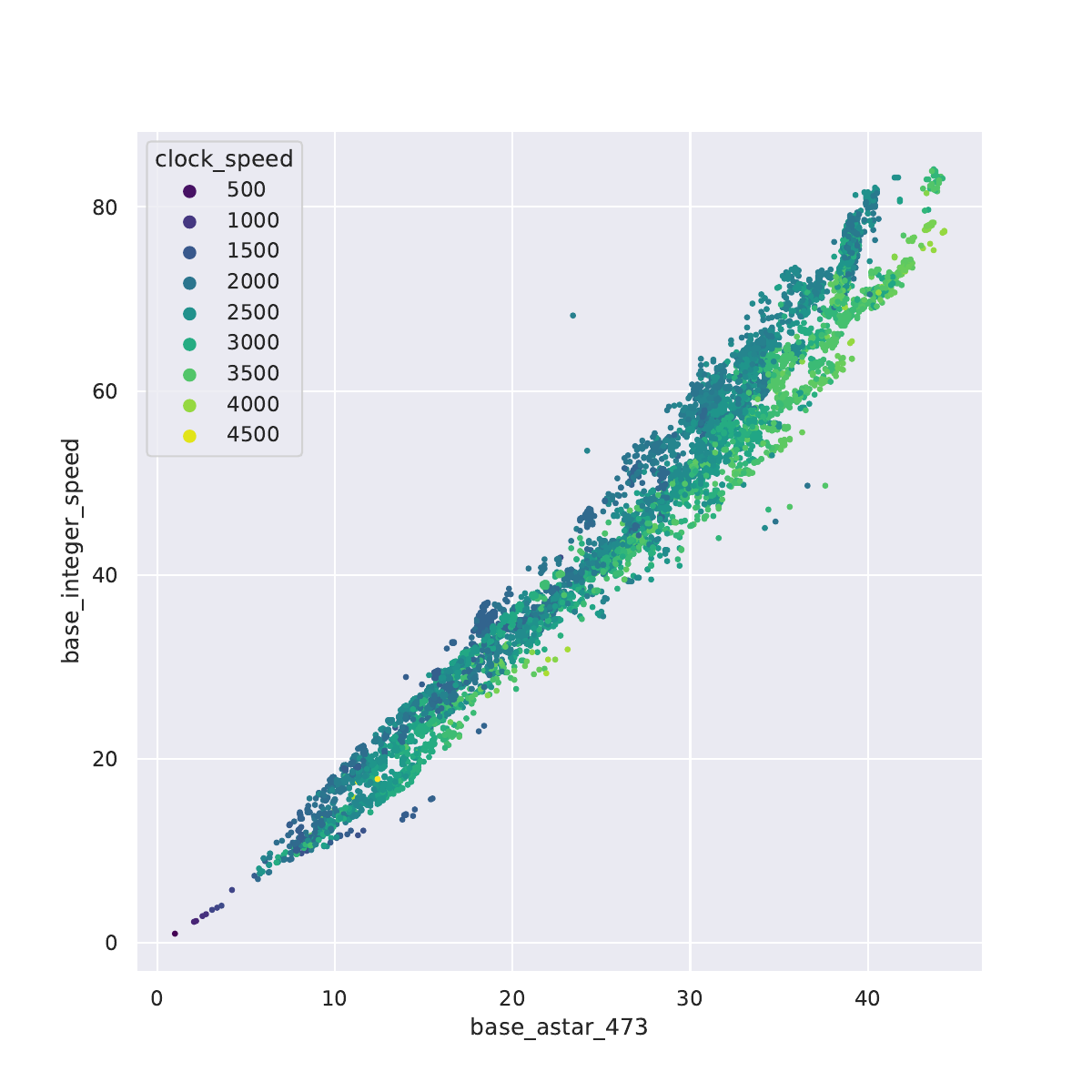}    &  \includegraphics[width = 0.3\linewidth]{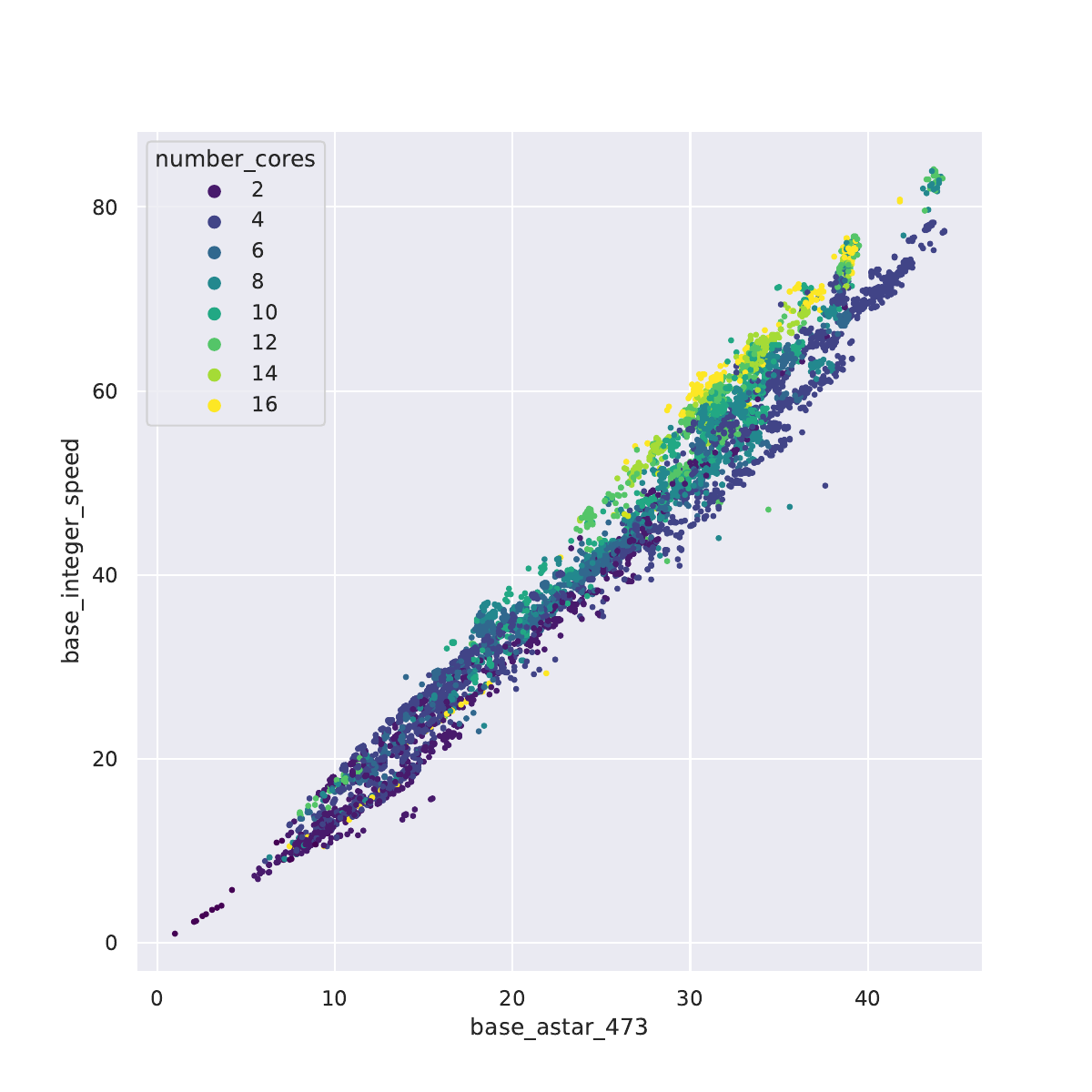}   \\
     (a) Frequency (MHz) & (b) Number of Cores\\
      \includegraphics[width = 0.3\linewidth]{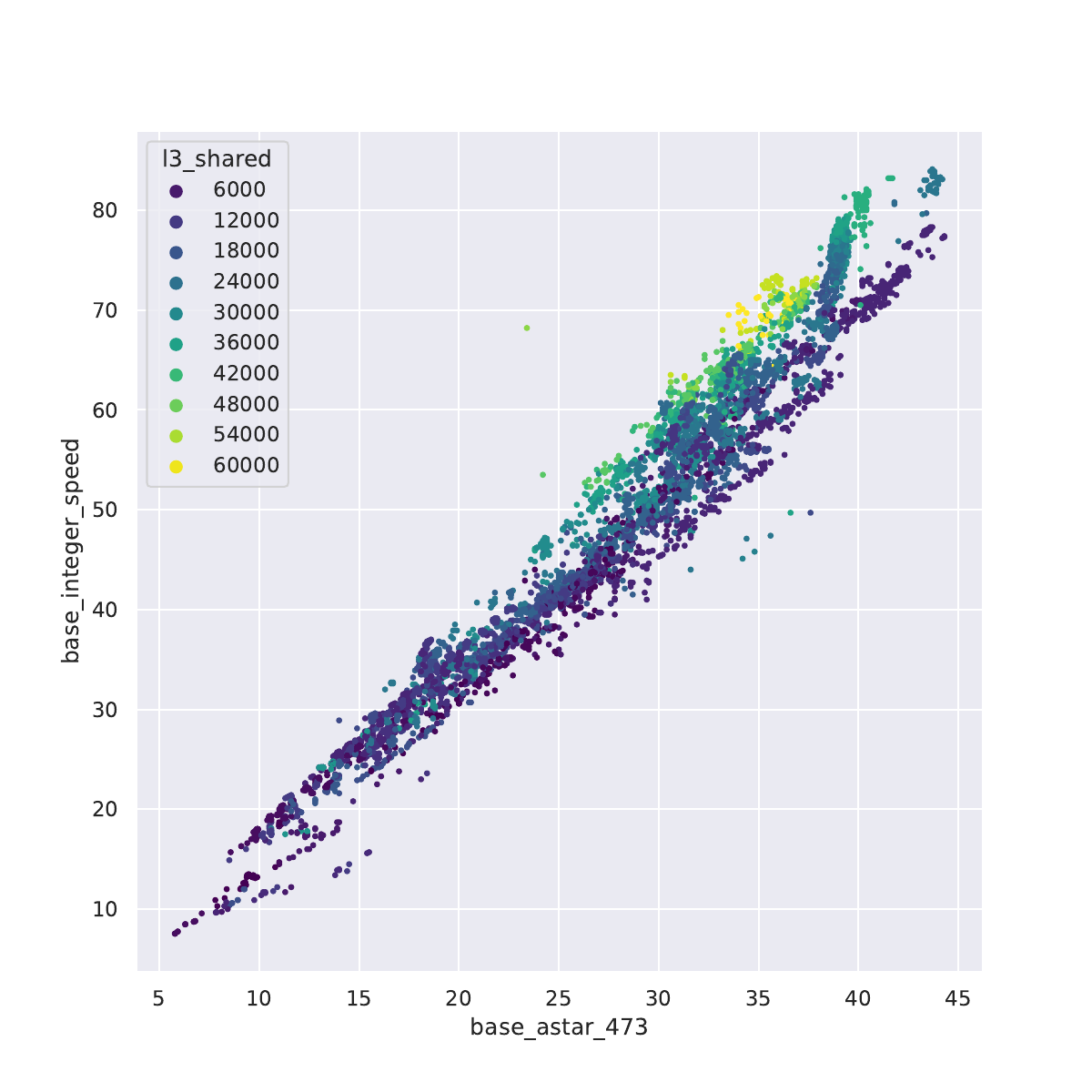}    &  \includegraphics[width = 0.3\linewidth]{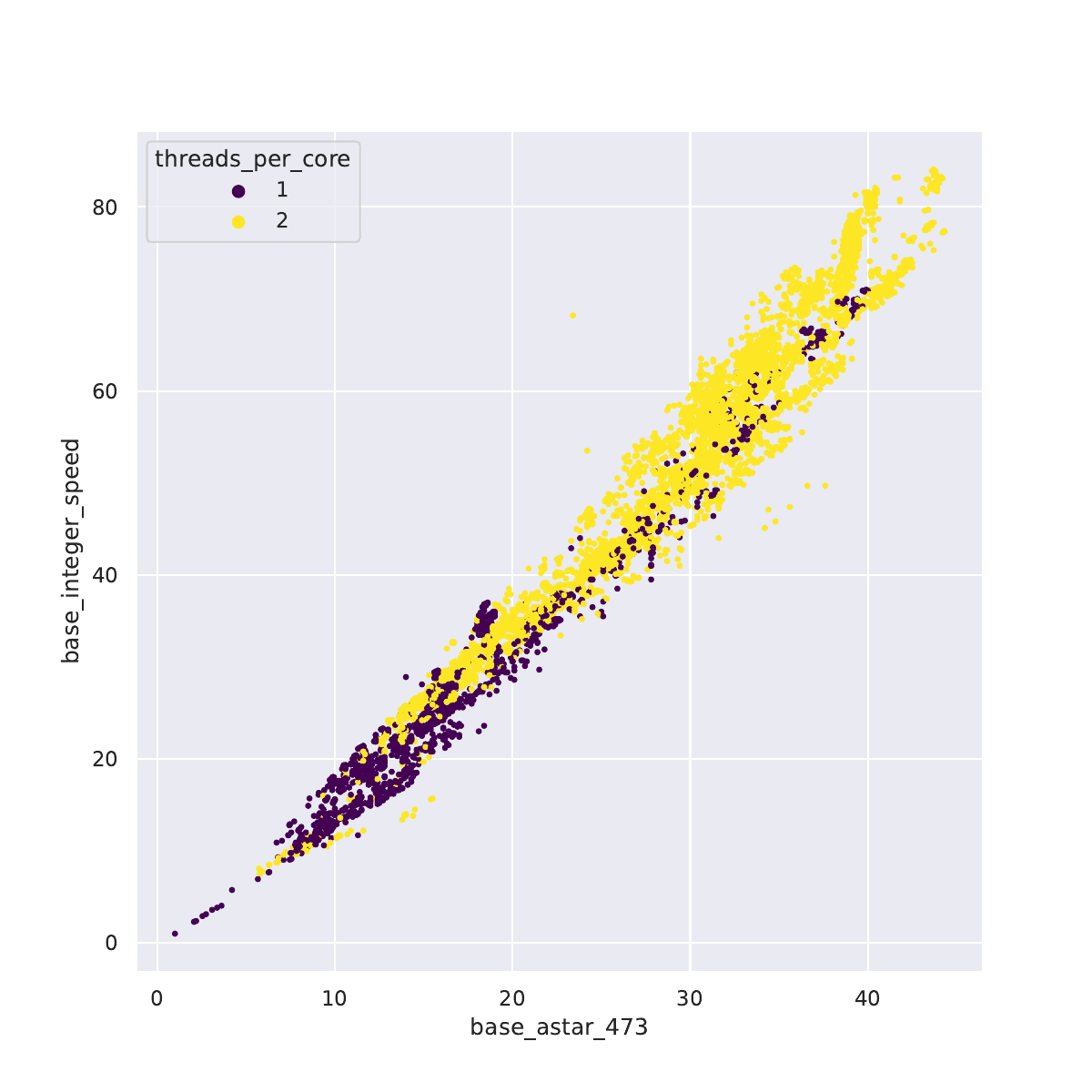}   \\
      (c) L3 Cache Size (KB) & (d) Threads per Core
\end{tabular}
    \caption{The system factors's impact on the relationship between base integer speed and astar. For machines with the same astar score, larger L3 cache sizes correlate with larger overall scores.}
    \label{fig:sys_over_astart}
\end{figure}

System factors also play an important role when understanding the relationship between the overall score and microbenchmarks. For SPEC 2006, we can take the microbenchmark ``astar\_473" as an example and look at its relationship with the overall score when the number of cores, frequency, transistor counts, and L3 cache size change. Figure~\ref{fig:sys_over_astart} (on page 7 for double-column figure) shows the relationship between astar and the overall score under different system factor settings. We can see that when the number of cores and threads are large, the records tend to have higher scores and move to the top right on the plots. For machines with the same astar score, larger L3 cache sizes correlate with larger overall scores.

\subsection{Libquantum}
\begin{figure}
    \centering
\begin{tabular}{cc}
 \includegraphics[width = 0.31\linewidth]{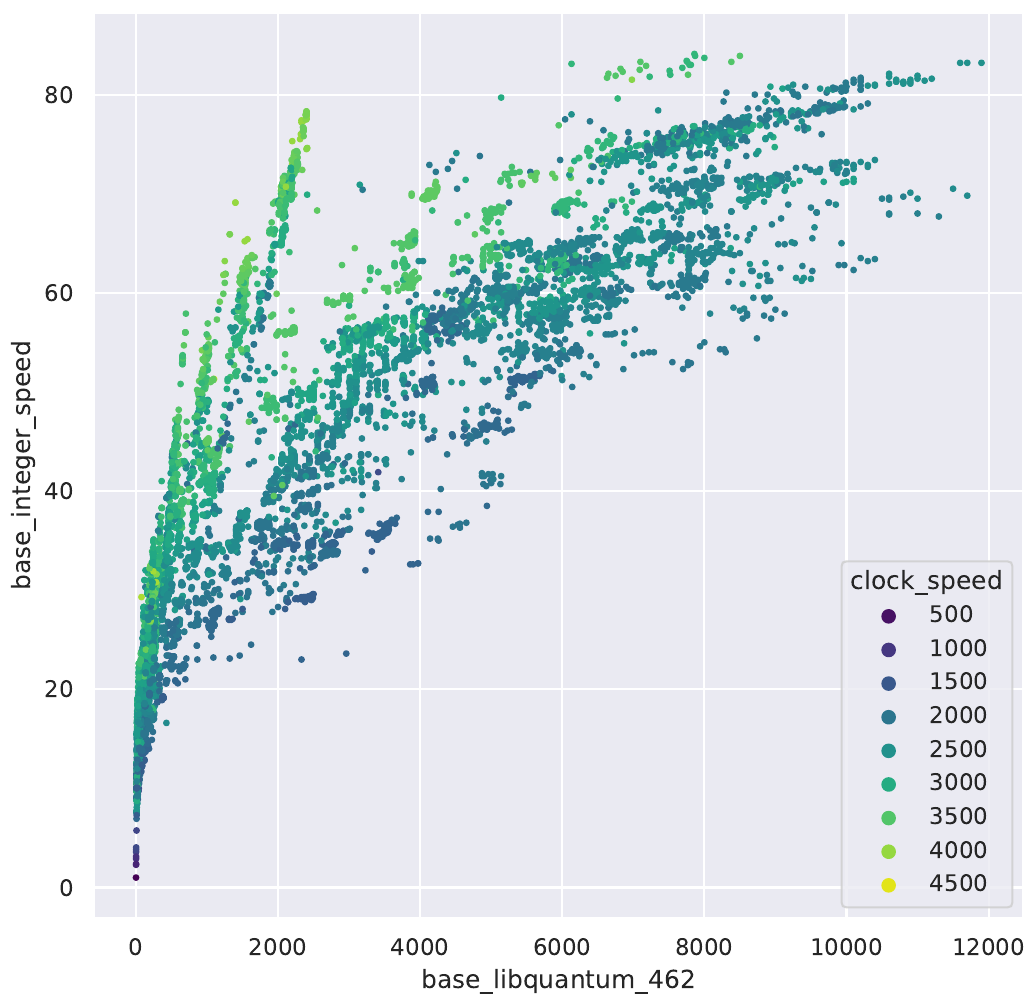}    &  \includegraphics[width = 0.31\linewidth]{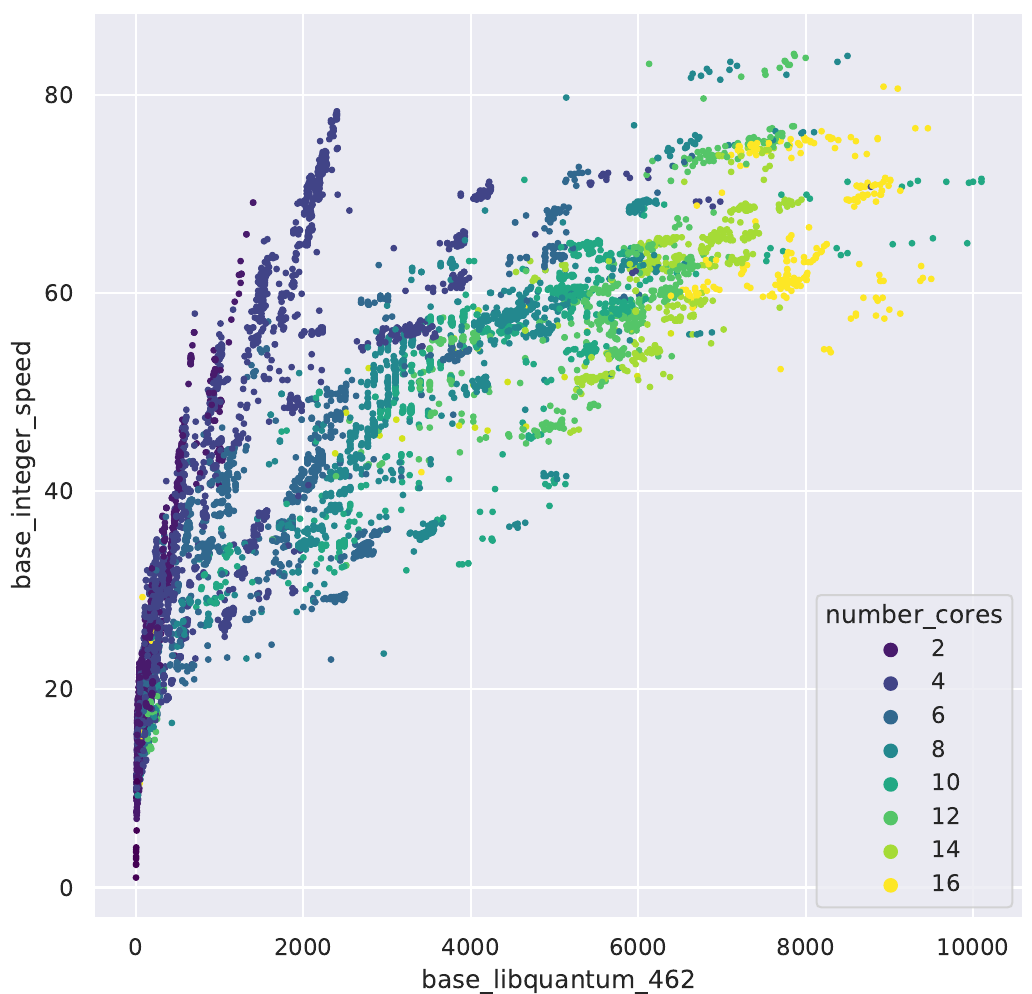}   \\
     (a) Frequency (MHz) & (b) Number of Cores\\
      \includegraphics[width = 0.31\linewidth]{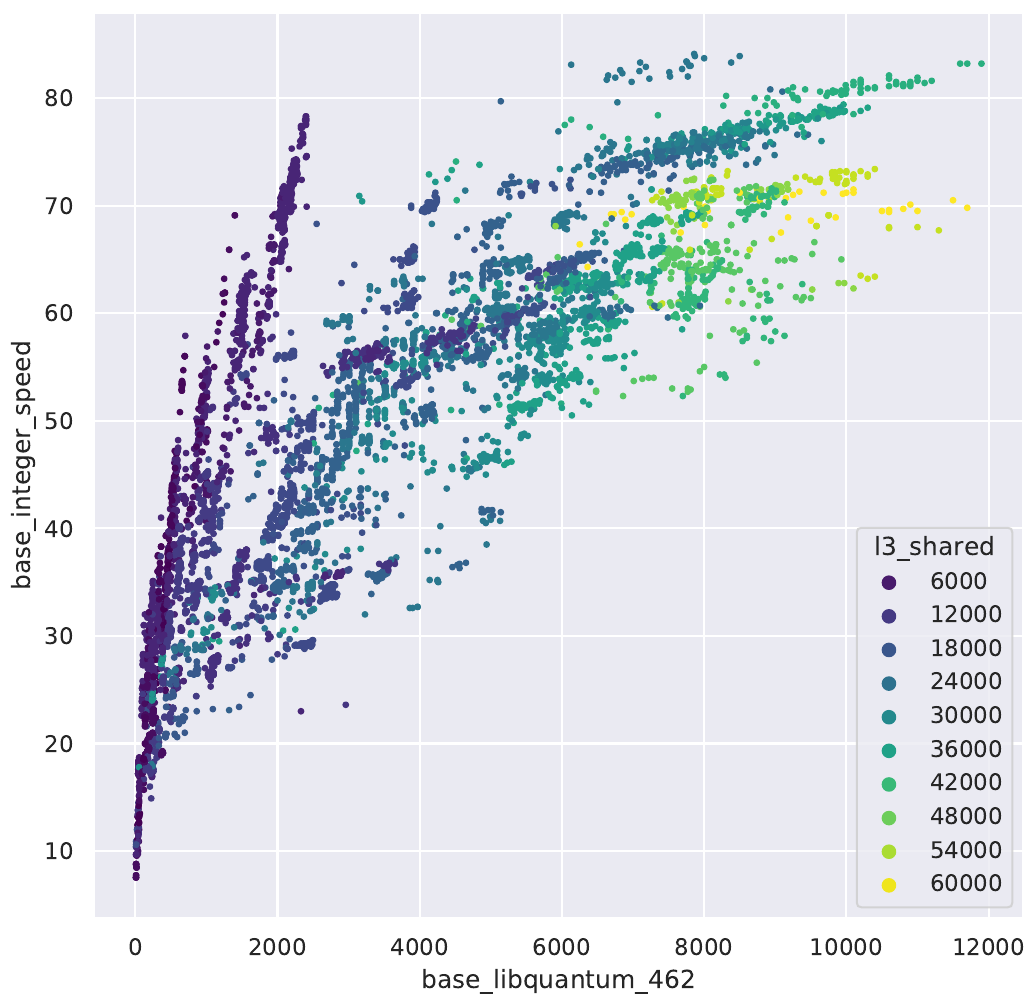}    &  \includegraphics[width = 0.31\linewidth]{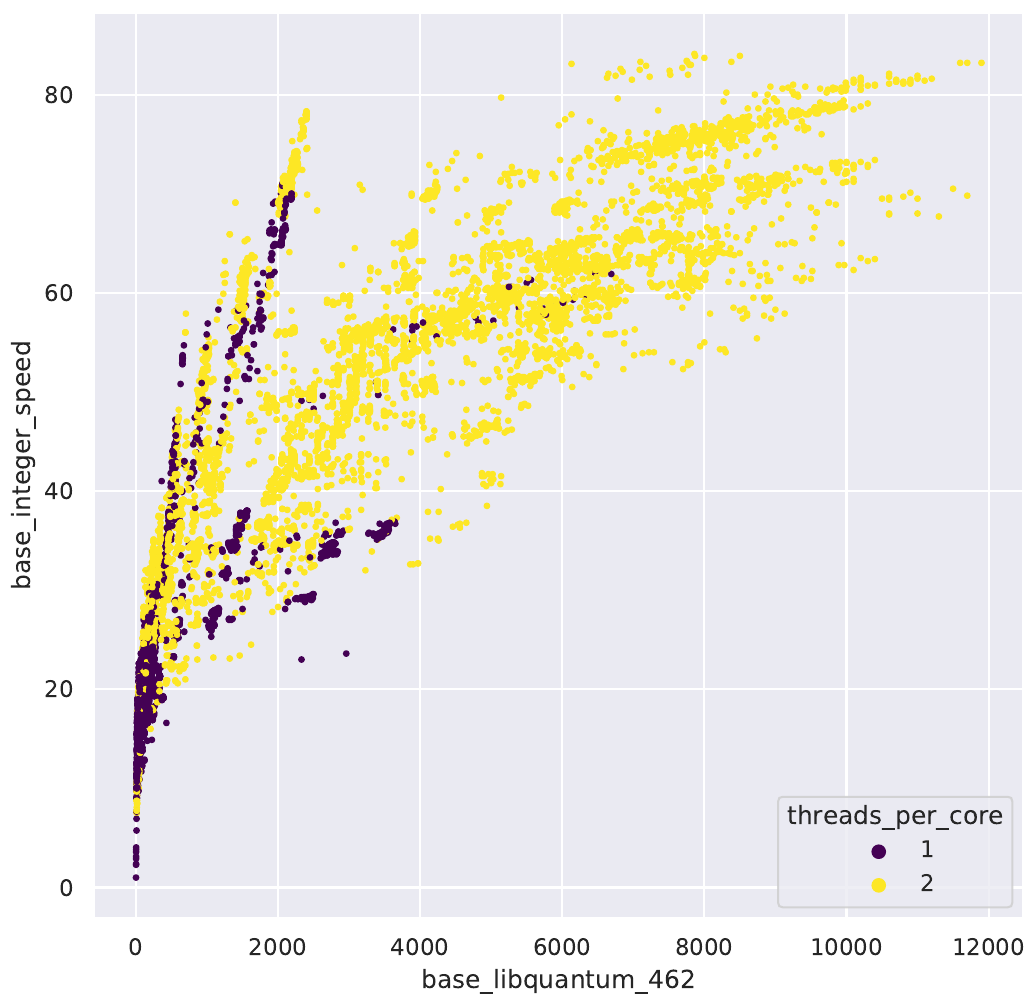}   \\
      (c) L3 Cache Size (KB) & (d) Threads per Core
\end{tabular}
    \caption{The system factors impact on the relationship between base integer speed and libquantum.}
    \label{fig:sys_over_lib}
\end{figure}

Among all the microbenchmarks, we notice that libquantum in SPEC 2006 is a leading factor in the base integer speed but does not influence the score ranking greatly. That means the changes in libquantum can have a great impact on the overall score but machines with large libquantum scores do not necessarily have top ranks. From visualizations, we observe that libquantum has higher variability than other microbenchmarks and is more sensitive to the number of cores. The microbenchmark libquantum has a very different pattern compared to other microbenchmarks. This microbenchmark has an inner for loop that can be optimized~\cite{stanford2012}. The removal of libquantum in SPEC 2017 indicates that it is desirable to let the score range and variability of microbenchmark scores be consistent when constructing benchmarks. As we can see from Figure~\ref{fig:sys_over_lib} (on page 8 for double-column figure), its relationship with base integer speed has a higher variance compared with astar in Figure~\ref{fig:sys_over_astart}. Its absolute score range is also much larger. In Figure~\ref{fig:sys_over_lib}, it shows that with a larger frequency, the increase in libquantum has a larger increase in base integer speed. Records with two threads per core have a larger variance compared to one thread per core. More visualizations are included in Figure~\ref{fig:3dlib}. This analysis also shows that it is desirable to let the score range and variability of microbenchmarks be consistent.

\subsection{Separate the Effect of System Factors}\label{sec:reg.sys}

Besides visualizing the impact of system factors, we also quantify and separate the effect of system factors on the overall speed score. We construct a linear regression to quantify the impact of core counts and the auto parallel on base integer speed. We assume there is a linear relationship between the normalized overall score on log scale and system factors. To reduce the co-linearity in the model, we did not include frequency in the model. In particular, the regression model is,
\begin{align*}
\text{Log Base Integer Speed} &= \beta_0 + \beta_1\,\text{core counts}+ \beta_2\,\text{auto\_parallel}.
\end{align*}
From Table~\ref{tbl:sys.overall.speed}, we can see that with the auto parallel flag on, the expectation of log base integer speed is 2.20 higher than without auto parallel system setting.

\begin{table}
\caption{Summary of regression for base integer speed versus system factors.}
\label{tbl:sys.overall.speed}

\begin{center}
\begin{tabular}{l|rrrr|rr}
\hline
\hline
Parameter  & Coef. & SE & $t$-val & $p$-val & \multicolumn{2}{c}{95\% CI}  \\
\hline
Intercept         &      $-$0.6959  &        0.013     &   $-$54.814  &         0.000        &       $-$0.721    &       $-$0.671     \\
No. of cores &       0.0281  &        0.001     &    47.184  &         0.000        &        0.027    &        0.029     \\
Auto-parallel      &       2.1994  &        0.015     &   151.380  &         0.000        &        2.171    &        2.228     \\
\hline
\hline
\end{tabular}
\end{center}

\end{table}

\subsection{Exploration on Lineage}
In the CSGenome repository, besides the benchmarks related information, we also have the lineage information \cite{sam2021thesis} about the processors. Processors' ancestors generations are gathered back to their commercial introduction. This information helps us to understand the relationships between processors and build a family tree of them. We are interested in whether the performance of different generations of processors that have the same origin are related.

\begin{figure}
    \centering
    \includegraphics[width = 0.7\linewidth]{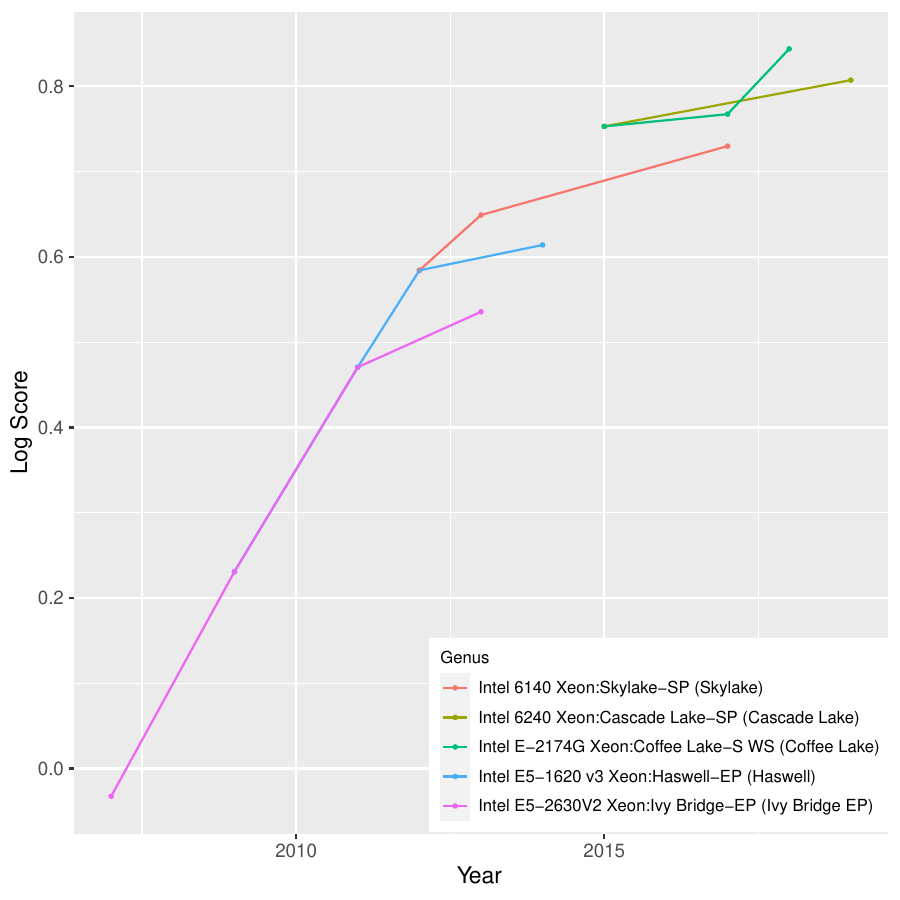}
    \vspace{-3ex}
    \caption{Log base integer score for 4 genus in SPEC 2017.}
    \label{fig:lineage.plot}
\end{figure}

We pick 5 represented processors in SPEC 2017 and use the lineage API to collect their ancestors' genus information up to 3 generations. A genus contains one or more processor models and one or more genera fall under a single microarchitecture. Then for each processor's genus and its ancestor genus, we compute their averaged performance scores and visualize these scores versus time. Figure~\ref{fig:lineage.plot} shows the averaged scores for each of the 5 representative genus and their ancestor branches in SPEC 2017. We can see that for each branch, recent generations have higher averaged performance. Among different branches, higher ancestor performance tends to have higher current generation performance. Besides, the correlations of lag-1 averaged performance score is 0.978. This indicates that lineage information is useful for us to understand future generations' performance.

\section{Prediction of Future Performance}\label{sec:prediction}

Another goal of this study is to predict future machine performance based on our data analysis results in Section~\ref{sec:results}. We provide both mean trend and individual computers' score predictions.

\subsection{Prediction for Mean Trend}\label{sec:moore.fit}
Many statistical and machine learning approaches can be adopted for mean trend prediction. For example, the ESTIMA tool presented in \cite{chatzopoulos2016estima} provides a flexible way to predict application performance trends. In this paper, we propose a nonlinear regression to capture the benchmark mean trend based on the observation in Figure~\ref{fig:integer.speed.norm2}.

\subsubsection{Nonlinear Regression and Normal Approximation}
Denote the overall performance score as $y_t$ and the time from baseline as $t$. The time unit is in months. Suppose the baseline score is $y_0$, as we observed in Figure~\ref{fig:integer.speed.norm2}, the log of the overall score appears to be a power function of time, indicating the period of performance doubling increases. Therefore, in this paper, we model the log of the overall score as
\begin{equation}\label{eq:moores.law}
\log(y_t) = f(t;\thetavec) + \epsilon_t = \alpha t^{\beta} + \gamma + \epsilon_t,
\end{equation}
where $\epsilon_t$ follows a normal distribution $\N(0, \sigma^2).$ Denote $\thetavec = (\alpha, \beta,
\gamma)'$ as the parameter vector. We use the maximum likelihood estimates (MLE) to conduct the parameter estimation in \eqref{eq:moores.law}. Then with the estimated parameters, we can make predictions of the performance score in the following years as well as provide prediction interval.

The variance of the log overall score is computed as,
$\var[\log(\yhat_t)] = f_{\thetavec}(t;\thetavechat)' \var(\thetavechat) f_{\thetavec}(t;\thetavechat),$
where $f_{\thetavec}(t;\thetavec)$ denotes the derivatives of $f(t;\thetavec)$ with respect to $\thetavec$. Then for a future time $t$, the point prediction is
$\log(\yhat_t) = \alphahat t^{\betahat} + \gammahat,$
and the 95\% prediction interval is
$$\log(\yhat) \pm z_{0.025} \sqrt{\var[\log(\yhat)] + \widehat{\sigma}^2}\,.$$

\subsubsection{Mean Trend Prediction Results}
\begin{figure}
    \centering
    \includegraphics[width =\linewidth]{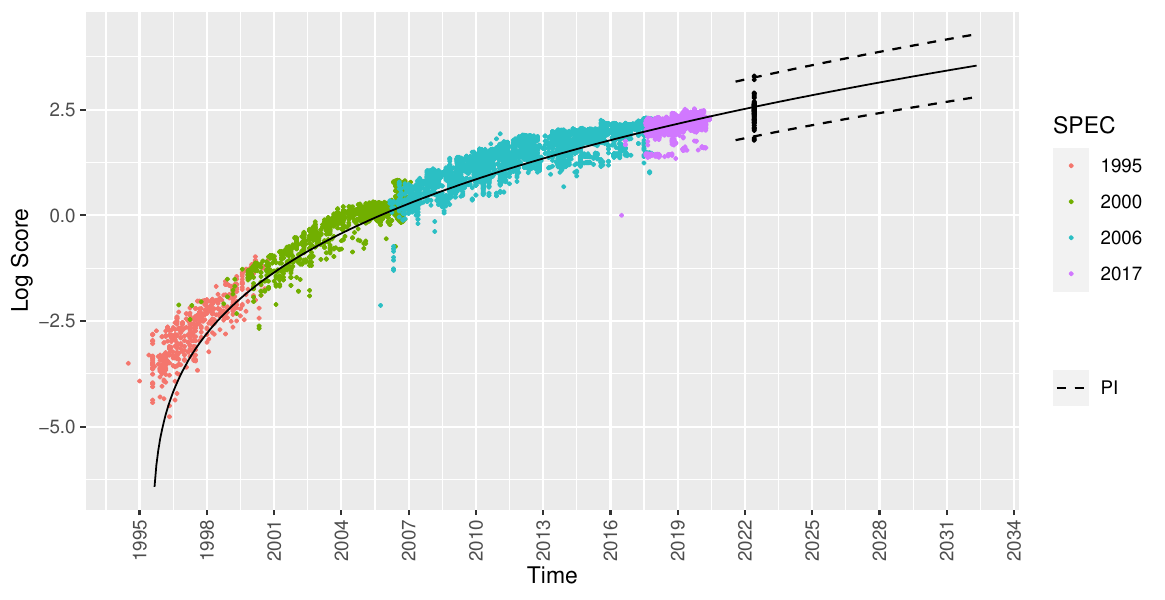}
    \caption{Mean trend fitting and prediction for base integer speed, together with individual predictions at a future time. The solid black curve represents the fitted mean trend and the dashed black lines are 95\% prediction bounds at future time. The black dots are predictions for the future score at 2022-06-01 based on (\ref{eq:pred.moores.sys}).}
    \label{fig:overall.ind.trend}
\end{figure}

For the integer speed, we obtain the estimates as $\thetavechat=(2.69, 0.25, -9.14)'$. By plugging in the MLE of the parameters in (\ref{eq:moores.law}), the fitted mean trend is
\begin{align}\label{eqn:estimated.moore.law}
\log(y) = 2.69 t^{0.25} -9.14.
\end{align}
The fitted mean trend and the prediction of the overall score for the future 100 months are shown in Figure~\ref{fig:overall.ind.trend}. The dashed line presents the 95\% prediction interval for the future prediction. As shown in Table~\ref{tbl:double}, we can see that for base integer speed, doubling the performance score needs to take a longer period. The baseline time is set to 1995-08-01, when the records start.

\begin{table}
\caption{Doubling time (in months) for the integer speed.}\label{tbl:double}
\begin{center}
\begin{tabular}{cc||cc||cc}
\hline\hline
 Date &  Time & Date &  Time & Date &  Time \\\hline
1996-04-01 &  0 & 1999-12-01 &  17 &  2010-11-01 &  45 \\
1996-10-01 &  6 & 2001-10-01 &  23 &  2015-05-01 &  55 \\
1997-07-01 &  9 & 2004-03-01 &  29 &  2020-10-01 &  66 \\
1998-08-01 & 13 &2007-03-01  &  36 &  2027-04-01 &  78 \\ \hline\hline
\end{tabular}
\end{center}
\end{table}

\subsection{Benchmark Prediction for Individual Computers}\label{sec:individual.pred}

Section~\ref{sec:moore.fit} provides fitting and prediction of the performance for the mean curve. That is the averaged performance over the years. That is because when we make predictions for the future, we usually do not know about the computers' hardware setting, so we can only make use of the time point as a predictor to provide predictions for scores on average. However, with the CSGenome database, we can also make predictions on individual computers with the system factors. Denote the system settings for a computer as $\xvec$, we assume the performance score of this computer at time $t$ can be modeled by two components: one is the average performance at time $t$ and the other is the fluctuation caused by this computer's system setting. So the performance score can be modeled as:
\begin{equation}\label{eq:moores.law.sys}
\log[y_t(\xvec)] = f(t;\thetavechat) +  \epsilon_t(\xvec)=\alphahat t^{\betahat} + \gammahat + \epsilon_t(\xvec),
\end{equation}
where the first component is the fitted mean trend as obtained in Section \ref{sec:moore.fit} and the second component is a random fluctuation that depends on $\xvec$.
Note that the difference between models~\eqref{eq:moores.law} and \eqref{eq:moores.law.sys} is that \eqref{eq:moores.law} assumes a constant noise variance for all different machines at time $t$, while \eqref{eq:moores.law.sys} assumes the noise term changes with different machine's system settings. Since the fitted mean trend is already obtained in Section \ref{sec:moore.fit}, the main goal of this section is to build a predictive model for $\epsilon_t(\xvec)$. Based on the sensitivity analysis in Section~\ref{sec:sensitivity.sys}, we use the hardware factors core counts, frequency, L3 cache size, and threads in SPEC 2017 as predictors to predict future $\epsilon_t(\xvec)$. Although auto-parallel is also an important factor as shown in Section~\ref{sec:reg.sys}, in SPEC 2017 records, 99.93\% of the machines consider auto-parallel when running SPEC speed benchmarks. So auto-parallel can be considered as a common feature for recent machines and becomes a constant variable if included in the model. Thus, we do not use it as a predictor for the future.

Before constructing the predictive models for $\epsilon_t(\xvec)$, there are a few preparation steps. First, we need to predict future hardware systems. Secondly, we need to exclude impossible hardware configurations based on reasonable extrapolation of existing processor data. Then based on the remaining reasonable configurations, we can build predictive models to predict the SPEC score.

\subsubsection{Quantile Regression for Hardware Prediction}\label{sec:qreg.sys}
\begin{figure}
    \centering
    \includegraphics[width = \linewidth]{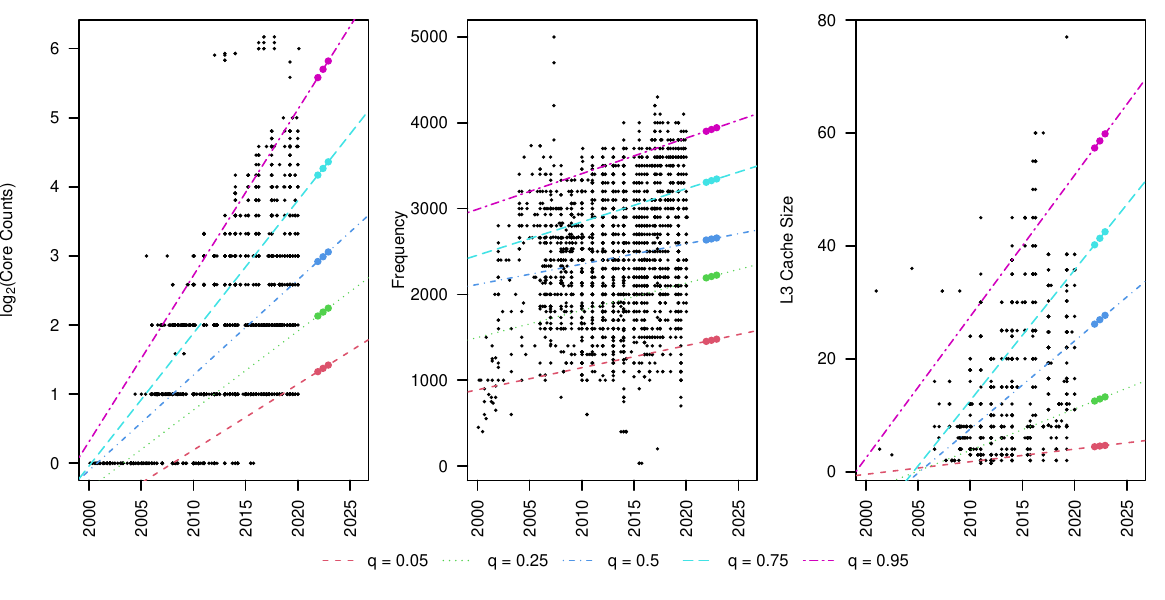}
    \vspace{-2ex}
    \caption{Plot of hardware factors over years and fitted quantile regression line and predictions of 0.05, 0.25, 0.5, 0.75, 0.95 quantile of core counts, frequency (MHz), and L3 cache size (MB) in three future time points (2021-12-01, 2022-06-01, and 2022-12-01). The core counts is on the $\log_2$ scale.}
    \label{fig:future.hw}
\end{figure}

We want to present some hardware scenarios in the future and make predictions of the corresponding machine's performance score. One key step is to obtain reasonable hardware settings for future computers. As shown in Figure~\ref{fig:future.hw}, starting around 2000, the trend of core counts, frequency, and L3 cache size versus time is not obvious and the variance becomes large. Therefore, when we make predictions for the future hardware settings, we want to capture both the trend and the dispersion of data. Quantile regression can model the quantile of hardware setting given the time, which allows us to understand both the variance and trend of hardware settings over time.

\subsubsection{Feasible Hardware Configurations}\label{sec:possible.configs}
\begin{figure}
    \centering
    \includegraphics[width = \linewidth]{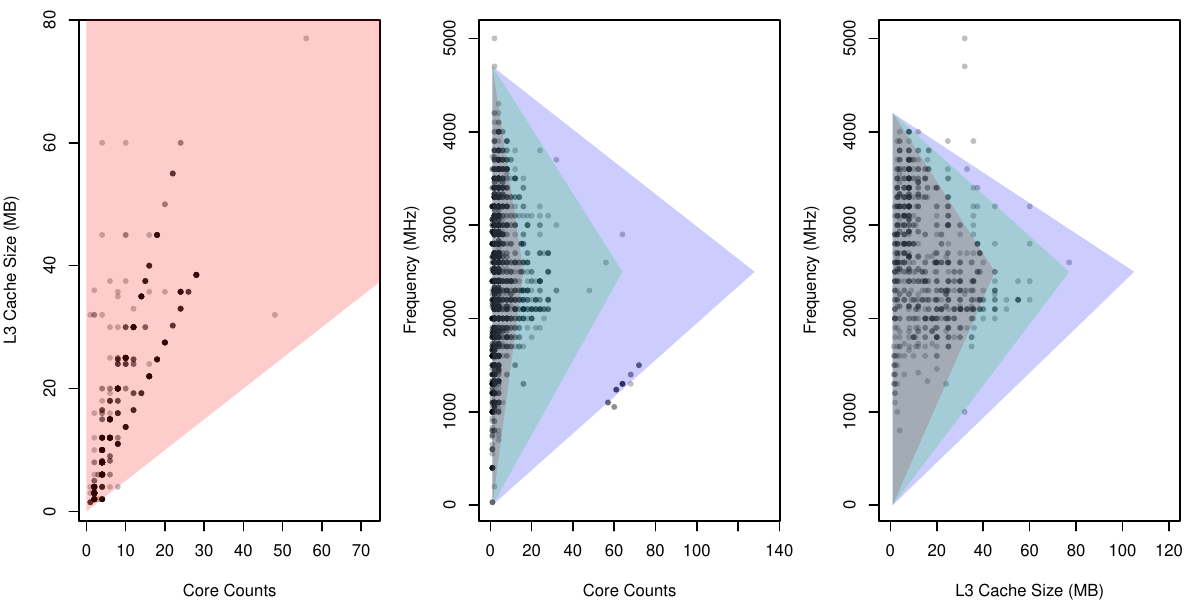}
    \vspace{-3ex}
    \caption{Extrapolation of the possible hardware regions in the future. The black points are historical data. In the left panel, the pink region represents the possible region of L3 cache size and core counts based on all historical data. In the middle and right panels, the grey area represents the year between 2000-2015. The blue area represents the year 2016-2020. The purple is the predicted area for 2021-2025.}
    \label{fig:feasible.config2}
\end{figure}

In our CSGenome data repository, we collected all related processor's information, which allows us to understand what configurations are feasible. From our processor data, since we are interested in the pattern between system factors, we focus on a subset with non-missing system factors. Within this subset, threads per core has two possible values, 1 and 2. So we only focus on the remaining three system factors, core counts, frequency, and L3 cache size. Figure~\ref{fig:feasible.config2} also presents the pairwise scatter plots of these three factors. We can see that for frequency and core counts, the region that has the most processor's data is roughly a triangular shape. That means as the core counts increase, the processors tend to have middle range of frequency. A similar pattern is found for the relationship between frequency and L3 cache size. For the relationship between the L3 cache size and core counts, based on the historical data, we found that the minimum ratio between them is 0.5 MB per core. So we consider those future processors with L3 cache size and core count ratio smaller than 0.5 MB per core are unlikely to appear.

Based on the above patterns, we make an extrapolation of the future possible regions of hardware configurations, which is shown in Figure~\ref{fig:feasible.config2}. In this figure, the shaded areas are the possible regions. The black points are historical data. The grey area represents the years between 2000-2015. The blue area represents the years 2016-2020. The purple is the predicted area for 2021-2025.

\subsubsection{Models for SPEC Score Prediction}
With these predicted reasonable future system factors, we can build a predictive model to forecast individual computer's benchmark scores. In this paper, we use the Gaussian process (GP) model to predict the normalized performance score for individual computers.

Suppose at time $t$ there is an individual computer with system factors $\xvec$, we can denote our response, the ``observed" noise as $\yeps_t = \epsilon_t(\xvec) = \log(y_t) - f(t;\thetavechat)$. Let $\yepsvec = \left\{\yeps_1, \dots, \yeps_n \right\}$ be the $n$ observations at the system factors collections $\Xvec = \left\{\xvec_1, \dots, \xvec_n \right\}$. The GP model assumes every finite linear combination of the observed data follow a multivariate normal (MVN) distribution. The mean function and the covariance function together determine a unique GP model. The mean function $\mu(\xvec)$ in the GP model provides the location parameter information depending on $\xvec$. The covariance function $C(\xvec, \xvec')$, which is also referred to as the kernel function, describes the correlations between the performance of machines with system factors $\xvec$ and $\xvec'$ respectively. The Gaussian kernel $C(\xvec, \xvec') = \exp[-{(\xvec - \xvec')^2}/{\theta}]$ and Mat\'ern kernels are common kernel functions.

\begin{figure}
    \centering
    \includegraphics[width = 0.82\linewidth]{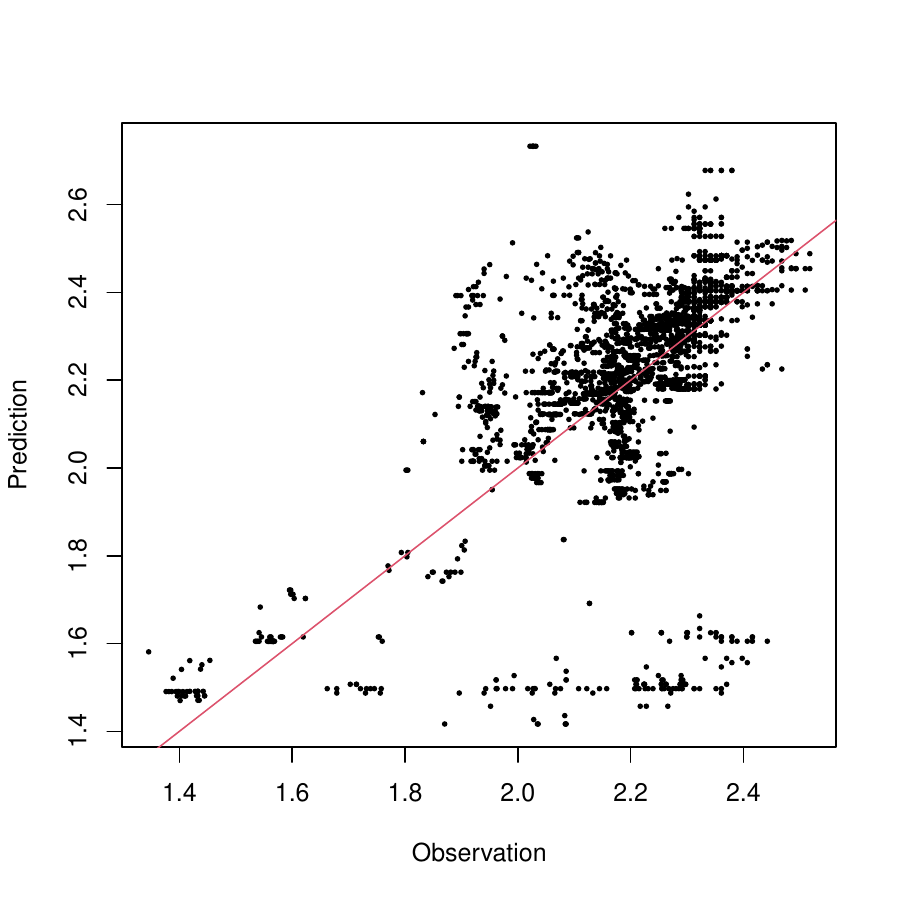}
    \vspace{-5ex}
    \caption{The GP predicted score versus observed score of the later 80\% SPEC 2017 machines. The red line is the 45 degree diagonal line.}
    \label{fig:gp.predict}
\end{figure}

In the GP model, the $n$ observed response is considered to follow MVN, that is $\yepsvec \sim \textrm{N}_n(0, \tau^2 \Cvec_n)$, where $\tau^2$ is the scale parameter and $\Cvec = [C(\xvec_i, \xvec_j)]_{n \times n}$ is the covariance matrix. Then for a new system configuration $\xvec$, the prediction of its performance score $\yeps(\xvec)$ can be obtained by the conditional distribution $\yeps_t(\xvec)|\left\{\yepsvec, \Xvec \right\}$. Based on the conditional distribution of MVN, we have
\begin{equation}\label{eq:gp.prediction}
\yeps_t(\xvec)|\left\{\yepsvec, \Xvec \right\} \sim \textrm{N}[\mu(\xvec), \sigma^2(\xvec)].
\end{equation}
The mean is $\mu(\xvec) = C(\xvec, \Xvec)\Cvec^{-1}\yepsvec$ and variance is
$$\sigma^2(\xvec) = \tauhat^2[C(\xvec, \xvec) - C(\xvec, \Xvec)\Cvec^{-1}C(\Xvec,\xvec)],$$ where, $C(\xvec, \Xvec_n)$ is a $1 \times n$ matrix with elements $C(\xvec, \xvec_1), \dots, C(\xvec, \xvec_n)$. The R package \textit{laGP} \cite{laGPpackage} is used to estimate parameters and construct GP models. To validate the GP model's ability to predict future, we use past data to predict the known future in SPEC 2017. The early 20\% computers (2016-09-01 to 2018-01-01) are used to predict the scores of the later 80\% computers (2018-02-01 to 2020-04-01). The root mean squared error is 0.19 and the predictions versus observations is shown in Figure~\ref{fig:gp.predict}.

\subsection{Prediction for Future Scenarios}

Considering the system configurations as well as the time in the benchmark score prediction, now the score of a machine with configuration $\xvec$ at time $t$ is predicted as
\begin{equation}\label{eq:pred.moores.sys}
    \log[\yhat_t(\xvec)] = f(t;\thetavechat) +\yepshat_t(\xvec) =\alphahat t^{\betahat} + \gammahat + \epsilonhat_t(\xvec).
\end{equation}
And the prediction's corresponding variance is
$$\Var\{\log[\yhat_t(\xvec)]\} = \Var[f(t;\thetavechat)] + \sigmahat^2(\xvec),$$
where $\Var[f(t;\thetavechat)]$ is in Section \ref{sec:moore.fit} and $\sigmahat^2(\xvec)$ is obtained from the GP model.
With this model, we can predict for future computers' performance score, and also construct prediction intervals for future computers' performance.

Based on Sections \ref{sec:qreg.sys} and \ref{sec:possible.configs}, we can obtain various possible hardware configurations future time point $t$. What we are interested in is the performance of top, middle and low rank computers. Therefore, we construct prediction bounds for the computers with $q$th quantile ($q$ = 0.25, 0.5, 0.75, and 0.95) performance scores at time $t$ using Algorithm~1 as follows.

\vspace{2ex}
\noindent\textbf{Algorithm 1:}
\begin{enumerate}
    \item At a future time point $t$, predict the 0.25, 0.5, 0.75, 0.95 quantiles of core counts, frequency, L3 cache size as $x_{\text{cores}}, x_{\text{cache}}$, and $x_{\text{freq}}$ separately;
    \item Generate future potential configurations at $t$ by enumerating all combinations of $x_{\text{cores}}, x_{\text{cache}}$, and $x_{\text{freq}}$;
    \item Exclude impossible configurations based on the plot shown in Figure~\ref{fig:feasible.config2} and denote remaining possible configurations as $\Xvec_t$;
    \item Predict the future score for each configuration in $\Xvec_t$ using \eqref{eq:pred.moores.sys} and denote the prediction as $\log\left(\yvec_{t}\right)$;
    \item Find the $q$th quantile of $\log(\yvec_{t})$ as well as its corresponding configuration $\xvec_{t}^q \in \Xvec_t$;
    \item For configuration $\xvec_{t}^q$, denote its prediction as $\log[\yhat_t(\xvec_t^q)]$ and compute its variance $\Var\{\log[\yhat_t(\xvec_t^q)]\}$.
    \item The 95\% prediction interval for a future $q$th quantile performance score at time $t$ is $\log[\yhat_t(\xvec_t^q)] \pm z_{0.025}\sqrt{\Var\{\log[\yhat_t(\xvec_t^q)]\}}$.
\end{enumerate}
\vspace{3ex}

\begin{figure}
    \centering
    \includegraphics[width = \linewidth]{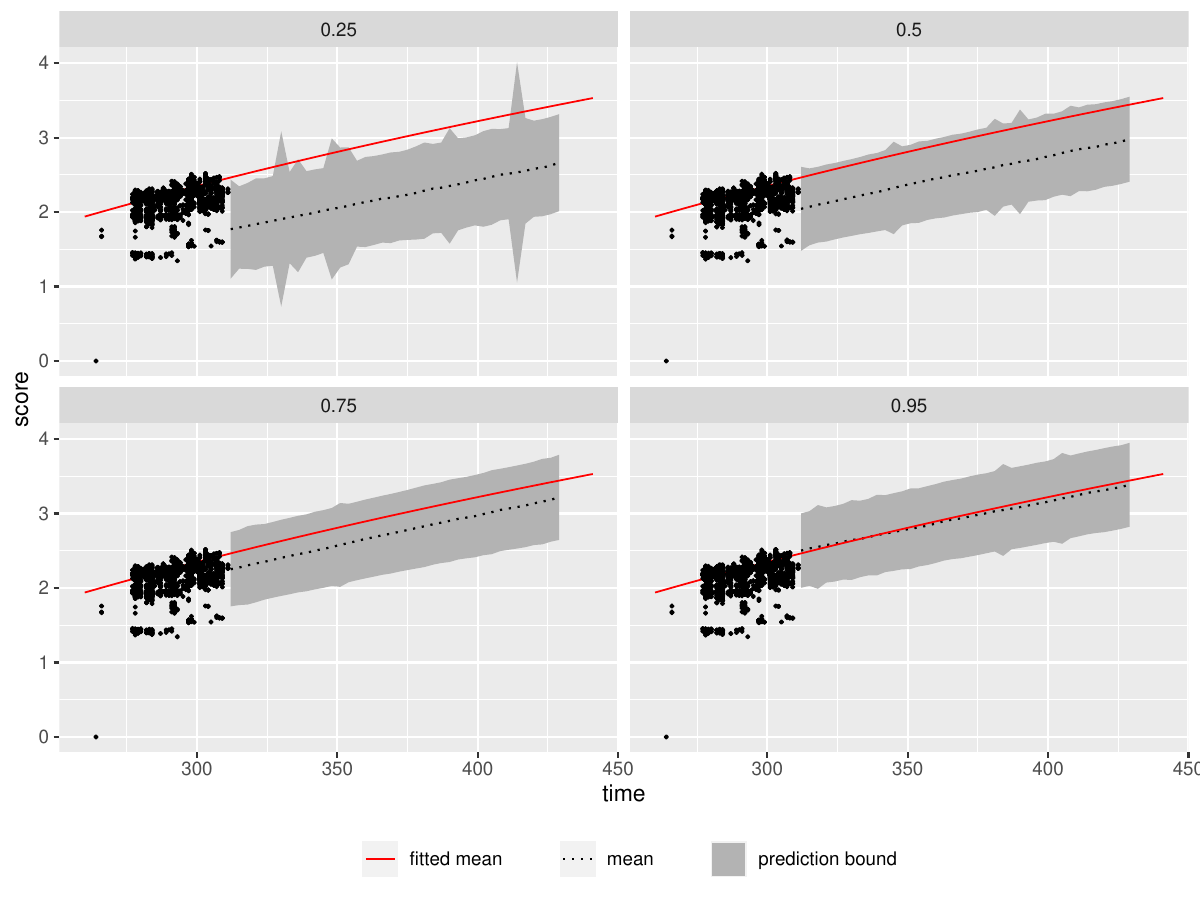}
    \vspace{-3ex}
    \caption{Prediction intervals for different future scenarios. The red solid line represents the mean trend fitting. The dotted black line represents the mean of each scenarios and the grey shaded areas are the 95\% prediction bounds for the corresponding scenario.}
    \label{fig:quanpi}
\end{figure}

Repeat Algorithm~1 for a sequence of future time points $t$, we can obtain a prediction bound for the future $q$th quantile performance scores.  The prediction bound of various future scenarios is shown in Figure~\ref{fig:quanpi}. The black dots are the real SPEC 2017 base integer speed that we used to build the GP model and the red line is fitted mean trend. In each subplot, the dotted black line presents the average $25\%, 50\%, 75\%$, and $95\%$ percentile of the score among all feasible future computers. The grey shaded bounds show the corresponding prediction bounds, respectively.

The predictive techniques developed in this paper can be used to answer the following questions in practice.

\begin{inparaitem}

\item For example, what characteristics will future systems need to keep up with previous trends? Our predicted mean trend depends on the model in \eqref{eqn:estimated.moore.law}, although in a diminishing pattern as the doubling time increases as shown in Table~\ref{tbl:double}.  The factors we measured, for example, the number of cores, clock frequency, or combination, need to maintain the trajectory we have enjoyed in the measurable past, as we modeled in Figure~\ref{fig:future.hw}.

\item The prediction results in Figure~\ref{fig:quanpi} provide insights on what kind of raw performance scores we should expect in the coming years. For example, using the mean and interval prediction for the 0.95 panel of Figure~\ref{fig:quanpi} (based on an optimistic prediction for technology development), we can see that the mean score (on log scale) will be around 3.37 (29.07 in original scale) in five years (2028-01-01) with a 95\% of interval from 3.20 to 3.59 (from 24.53 to 36.23 on the original scale) depending on how the technologies scale.

\item Based on a pessimistic prediction for technology development, the results in the 0.25 panel of Figure~\ref{fig:quanpi} show that the performance gains will not likely continue with the mean trend. To get us back on track,  consistent hardware improvements are needed, for example, on the number of cores and clock frequency. The 0.75 panel of Figure~\ref{fig:quanpi} gives us a clue as to the kind of technology improvement needed to continue with the current pattern on performance.

\end{inparaitem}

Although we present the prediction procedure using the SPEC base integer speed as an example, the framework of parameter estimation, uncertainty quantification, and predictions can be applied to new benchmark exploration. If the performance trend has a different pattern in the new benchmark datasets, the proposed method can be adapted by changing the function form in the model in \eqref{eq:moores.law}.

\section{Concluding Remarks} \label{conclusion}
In this paper, we have considered how best to track and analyze historical SPEC data represented by SPEC CPU integer speed. We selected this specific benchmark since it is one of the most widely used and historically thorough benchmarks for computer systems. This focus does limit the scope of our conclusions, but the methodology regarding normalizing benchmark scores, sensitivity analysis, and prediction framework could be applied to other benchmarks and computing performance datasets. We overcame several challenges along the way including determining the best available methods for normalization as the benchmark evolves. Additionally, we isolated the effects of individual codes and determined one code in particular (libquantum) had outsized influence over benchmark scores -- a fact that likely led to its removal as its use could have allowed participants to exploit the code for their own gain in rankings.

We used a growing, open-source database of computer specifications and lineage to study the impact of design decisions (e.g., core count, cache size) on performance over time. We confirmed some expectations: 1) multi-core processors begin to dominate as the effect of individual processor speed diminish; 2) SPEC CPU integer speed performance tracks with Moore's Law mostly; and 3) after 2000, the influence of several hardware traits (e.g., L3 size, core count) on performance becomes more murky and more difficult to point to a single contributor.

As for future designs, we discussed a methodology that determines how the base technologies studies will need to evolve to continue to track with Moore's Law. In the future, L3 cache sizes and core counts will likely have the most influence over future designs without disruptive change.

We want to point out that disruptive changes in hardware, such as AMD's 3D V-cache can impact our prediction potentially. At the moment that new technology advances emerge, precisely predicting the performance score is challenging if new data are too far from all other historical data. However, there have been other inflection points in past years, such as bigger caches, faster processors, and on-core GPUs that are already captured in the data and our prediction methodology is constructed with such evolved hardware data. As the database grows, the changes with new hardware will be captured eventually and the predictions will adapt modestly as well by re-applying the framework.

There are a few limitations in our work. In the analysis, we mainly focused on the hardware effect and integrated software effect into the results. However, software and compiler's development also contribute to the system's performance. In future studies, it is interesting to isolate and quantify software effects to further understand computers' performance evolution trend as well as Proebsting's Law. Besides, the analysis scope focuses on benchmark evolution on the system level. If the focus is extended to a broader scope of benchmarks not limited to CPU or system level and the trend of interest is too complicated to describe, other tools that automatically select function forms, like ESTIMA \cite{chatzopoulos2016estima} can be considered.

\section*{Acknowledgments}

The authors thank the editor, associate editor, and three referees, for their valuable comments that helped improve the paper significantly.
The authors acknowledge Advanced Research Computing at Virginia Tech for providing computational resources.
The work was supported by NSF CNS-1838271 and CNS-1939076 to Virginia Tech.

\appendix

\section{Overall Score and Microbenchmarks}
\begin{figure}
    \centering
    \begin{tabular}{c}
        \includegraphics[width = \linewidth]{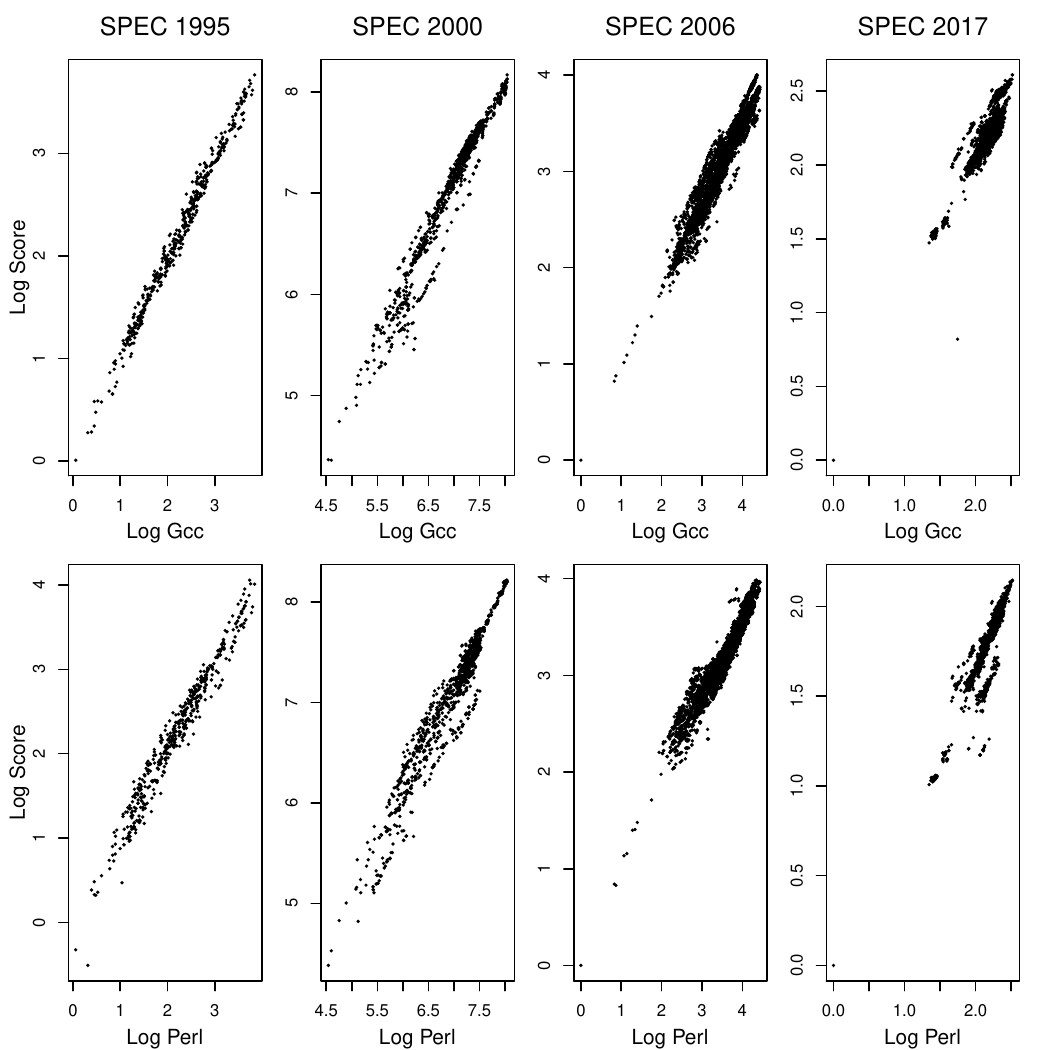} \\
    \end{tabular}

    \caption{Base integer speed versus gcc and perl for each SPEC suite with log transformation.}
    \label{fig:overall_micro_speed}
\end{figure}

Figure~\ref{fig:overall_micro_speed} visualizes base integer speed versus two microbenchmarks: gcc and perl for each SPEC suite, respectively. We can see that after log transformation, the log overall score and log microbenchmark scores are highly linearly correlated.

\section{More Visualizations on Libquantum}
\begin{figure}
\begin{center}
\includegraphics[width = 1.0\linewidth]{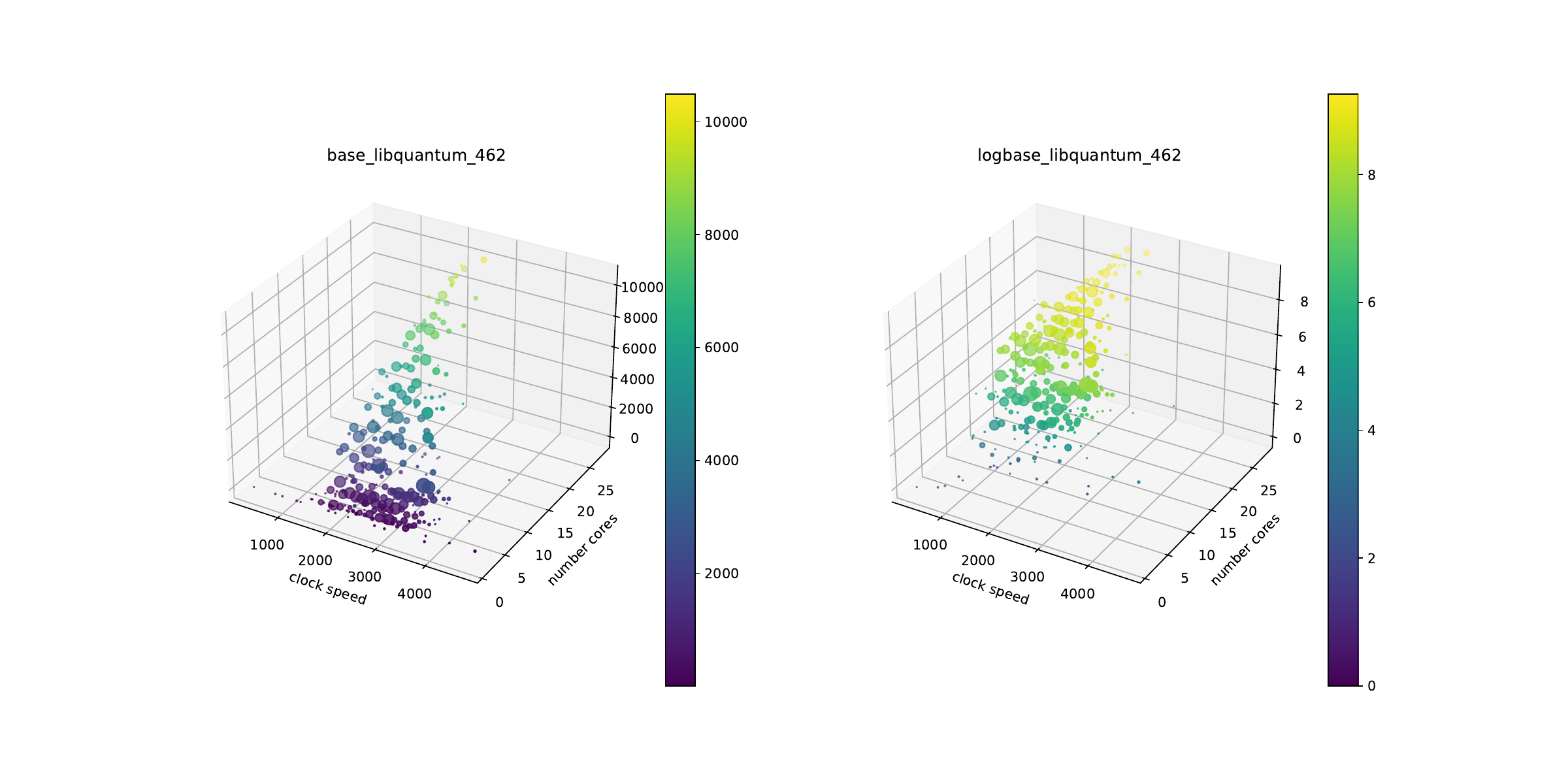}
\vspace{-3ex}
\caption{Frequency (MHz) and core counts impact on the libquantum (left) and log libquantum (right).}\label{fig:3dlib}
\end{center}
\end{figure}

Figure~\ref{fig:3dlib} visualizes the frequency and core counts influence to the ratio between libquantum and base integer speed. When the number of cores is small, libquantum changes a lot relatively (ratio of base integer speed to libquantum) and when the number of cores is small (with slow clock speed), libquantum (with clock speed) will affect overall metric more.

\begin{thebibliography}{10}
\providecommand{\url}[1]{#1}
\csname url@samestyle\endcsname
\providecommand{\newblock}{\relax}
\providecommand{\bibinfo}[2]{#2}
\providecommand{\BIBentrySTDinterwordspacing}{\spaceskip=0pt\relax}
\providecommand{\BIBentryALTinterwordstretchfactor}{4}
\providecommand{\BIBentryALTinterwordspacing}{\spaceskip=\fontdimen2\font plus
\BIBentryALTinterwordstretchfactor\fontdimen3\font minus
  \fontdimen4\font\relax}
\providecommand{\BIBforeignlanguage}[2]{{%
\expandafter\ifx\csname l@#1\endcsname\relax
\typeout{** WARNING: IEEEtran.bst: No hyphenation pattern has been}%
\typeout{** loaded for the language `#1'. Using the pattern for}%
\typeout{** the default language instead.}%
\else
\language=\csname l@#1\endcsname
\fi
#2}}
\providecommand{\BIBdecl}{\relax}
\BIBdecl

\bibitem{specgrowth}
J.~L. Henning, ``{SPEC} {CPU} suite growth: an historical perspective,''
  \emph{ACM SIGARCH Computer Architecture News}, vol.~35, no.~1, pp. 65--68,
  2007.

\bibitem{spec2017broaden}
R.~Panda, S.~Song, J.~Dean, and L.~K. John, ``Wait of a decade: Did {SPEC}
  {CPU} 2017 broaden the performance horizon?'' in \emph{2018 IEEE
  International Symposium on High Performance Computer Architecture}, ser. HPCA
  '18.\hskip 1em plus 0.5em minus 0.4em\relax Washington, DC: IEEE Computer
  Society, 2018, pp. 271--282.

\bibitem{specsamebottlenecks}
H.~Vandierendonck and K.~De~Bosschere, ``Many benchmarks stress the same
  bottlenecks,'' in \emph{Workshop on Computer Architecture Evaluation Using
  Commercial Workloads}, ser. CAECW '04.\hskip 1em plus 0.5em minus 0.4em\relax
  Washington, DC: IEEE Computer Society, 2004, pp. 57--64.

\bibitem{specprogramsimilarity}
A.~Phansalkar, A.~Joshi, L.~Eeckhout, and L.~John, ``Measuring program
  similarity: Experiments with {SPEC} {CPU} benchmark suites,'' in \emph{IEEE
  International Symposium on Performance Analysis of Systems and Software,
  2005}, ser. ISPASS '05.\hskip 1em plus 0.5em minus 0.4em\relax Washington,
  DC: IEEE Computer Society, 2005, pp. 10--20.

\bibitem{specredundancy}
A.~Phansalkar, A.~Joshi, and L.~K. John, ``Analysis of redundancy and
  application balance in the {SPEC} {CPU2006} benchmark suite,'' in
  \emph{Proceedings of the 34th Annual International Symposium on Computer
  Architecture}, ser. ISCA '07.\hskip 1em plus 0.5em minus 0.4em\relax New
  York, NY: Association for Computing Machinery, 2007, pp. 412--423.

\bibitem{spec2006subsetting}
------, ``Subsetting the {SPEC} {CPU2006} benchmark suite,'' \emph{SIGARCH
  Comput. Archit. News}, vol.~35, no.~1, pp. 69–--76, 2007.

\bibitem{stanford2012}
A.~Danowitz, K.~Kelley, J.~Mao, J.~P. Stevenson, and M.~Horowitz, ``{CPU} {DB}:
  Recording microprocessor history: With this open database, you can mine
  microprocessor trends over the past 40 years.'' \emph{ACM Queue}, vol.~10, p.
  10–27, 2012.

\bibitem{sam2021thesis}
\BIBentryALTinterwordspacing
S.~L. Furman, ``{iLORE}: Discovering a lineage of microprocessors,'' 2021,
  {M}aster's Thesis, Virginia Tech. [Online]. Available:
  \url{http://hdl.handle.net/10919/104071}
\BIBentrySTDinterwordspacing

\bibitem{speccpu}
SPEC, ``speccpu benchmarks,'' \url{https://www.spec.org/benchmarks.html}, 2021.

\bibitem{evolandevalspec}
J.~J. Dujmovic and I.~Dujmovic, ``Evolution and evaluation of {SPEC}
  benchmarks,'' \emph{SIGMETRICS Perform. Eval. Rev.}, vol.~26, no.~3, pp.
  2–--9, 1998.

\bibitem{nicolas2021thesis}
\BIBentryALTinterwordspacing
N.~R. Hardy, ``A data schema for aggregating disparate sources of computer
  system and benchmark information,'' 2021, {Ma}ster's Thesis, Virginia Tech.
  [Online]. Available: \url{http://hdl.handle.net/10919/103707}
\BIBentrySTDinterwordspacing

\bibitem{bs4}
B.~Soup, ``{Beautiful Soup},''
  \url{https://www.crummy.com/software/BeautifulSoup/bs4/doc/}, 2021.

\bibitem{Pandas}
Pandas, ``Pandas,'' \url{https://pandas.pydata.org/}, 2021.

\bibitem{top500lists}
TOP500, ``Top500 lists,'' \url{https://www.top500.org/}, 2021.

\bibitem{green500lists}
Green500, ``Green500 lists,'' \url{https://www.top500.org/lists/green500/},
  2023.

\bibitem{amdprocessorspecs}
AMD, ``{AMD} product specifications,''
  \url{https://www.amd.com/en/products/specifications/processors}, 2021.

\bibitem{arkintel}
Intel, ``{Intel} product specifications,''
  \url{https://ark.intel.com/content/www/us/en/ark.html}.

\bibitem{stanfordcpudbwebsite}
A.~Danowitz, ``Stanford {CPU} {DB},'' \url{http://cpudb.stanford.edu/}, 2014.

\bibitem{chatzopoulos2016estima}
G.~Chatzopoulos, A.~Dragojevi{\'c}, and R.~Guerraoui, ``Estima: Extrapolating
  scalability of in-memory applications,'' in \emph{Proceedings of the 21st ACM
  SIGPLAN Symposium on Principles and Practice of Parallel Programming}, 2016,
  pp. 1--11.

\bibitem{laGPpackage}
R.~B. Gramacy, ``{laGP}: Large-scale spatial modeling via local approximate
  {Gaussian} processes in {R},'' \emph{Journal of Statistical Software},
  vol.~72, no.~1, pp. 1--46, 2016, doi: 10.18637/jss.v072.i01.

\end{thebibliography}


\end{document}